\documentclass[11pt,a4paper]{article}
\usepackage[top=2.5cm,left=2cm,footskip=1cm,marginparwidth=0cm]{geometry}

\usepackage[utf8x]{inputenc}
\usepackage{subfloat}
\usepackage{lineno,hyperref}
\usepackage{lmodern,textcomp}
\usepackage{amsmath}
\usepackage[lf]{berenis}
\usepackage[LY1]{fontenc}

\usepackage[round]{natbib}

\usepackage{nameref,hyperref}

\usepackage{microtype}
\DisableLigatures[f]{encoding = *, family = * }

\raggedright
\usepackage{changepage}

\usepackage[aboveskip=1pt,labelfont=bf,labelsep=period,singlelinecheck=off]{caption}

\usepackage{lastpage,fancyhdr,graphicx}
\usepackage{epstopdf}
\usepackage{color}

\definecolor{Gray}{gray}{.25}

\usepackage{graphicx}

\usepackage{sidecap}

\usepackage{wrapfig}
\usepackage[pscoord]{eso-pic}
\usepackage[fulladjust]{marginnote}
\reversemarginpar

\begin{document}
\vspace*{0.35in}

\begin{flushleft}
{\Large
\textbf\newline{Systematic measurements of the night sky brightness at 26 locations \linebreak in Eastern Austria}
}
\newline
\\
Thomas Posch\textsuperscript{1,*},
Franz Binder\textsuperscript{1},
Johannes Puschnig\textsuperscript{2}
\\
\bigskip
\bf{1} Universit{\"a}t Wien, Institut für Astrophysik, T\"urkenschanzstraße 17, A-1180 Wien, Austria\\
\bf{2} Zentrum für Astronomie der Universität Heidelberg, Institut für Theoretische Astrophysik,
Albert-\"Uberle-Str.\ 2, D-69120 Heidelberg, Germany\\
\bigskip
* tel: +43 1 4277 53800, e-mail: thomas.posch@univie.ac.at

\end{flushleft}

\section*{Abstract}
We present an analysis of the zenithal night sky brightness (henceforth: NSB) measurements at 26 locations in Eastern Austria focussing on the years 2015--2016, both during clear and cloudy to overcast nights. All measurements have been performed with 'Sky Quality Meters' (SQMs). For some of the locations, simultaneous aerosol content measurements are available, such that we were able to find a correlation between light pollution and air pollution at those stations. For all locations, we examined the circalunar periodicity of the NSB, seasonal variations as well as long-term trends in the recorded light pollution. For several remote locations, a darkening of the night sky due to clouds by up to 1 magnitude is recorded -- indicating a very low level of light pollution --, while for the majority of the examined locations, a brightening of the night sky by up to a factor of 15 occurs due to clouds. We present suitable ways to plot and analyze huge long-term NSB datasets, such as mean-NSB histograms, circalunar, annual ('hourglass') and cumulative ('jellyfish') plots. We show that five of the examined locations reach sufficiently low levels of light pollution -- with NSB values down to 21.8 mag$_{SQM}$/arcsec$^{2}$ -- as to allow the establishment of dark sky reserves, even to the point of reaching the 'gold tier' defined by the International Dark Sky Association. Based on the 'hourglass' plots, we find a strong circalunar periodicity of the NSB in small towns and villages ($<$ 5.000 inhabitants), with amplitudes of  of up to 5 magnitudes. Using the 'jellyfish' plots, on the other hand, we demonstrate that the examined city skies brighten by up to 3 magnitudes under cloudy conditions, which strongly dominate in those cumulative data representations. Nocturnal gradients of the NSB of 0.0--0.14 mag$_{SQM}$/arcsec$^{2}$/hr are found. The long-term development of the night sky brightness was evaluated based on the 2012--17 data for one of our sites, possibly indicating a slight ($~$2\%) decrease of the mean zenithal NSB at the Vienna University Observatory.

\nolinenumbers

\section{Introduction}

{Humanity is facing a number of novel environmental problems resulting from the technological and economical advances of the last hundred years \citep{dunlap2012environmental}. One of these challenges is light pollution. Recently, a lot of efforts were devoted to research on this topic and its effects on ecosystems \citep{longcore2004ecological}, human health \citep{chepesiuk2009missing}, animals and plants. The rapid global growth of artificially lit outdoor areas, pinpointed by \citet{Kybae1701528}, shows the need for further monitoring and documentation of the current situation in terms of light intensities, skyglow and spectral composition of nocturnal illumination.} 

{As for the threats of light pollution to ecosystems, one example is the worldwide decline of pollinators \citep{potts2010global}, which play a key role in human food security. \citet{knop2017artificial} and \citet{macgregor2015pollination} recently pointed out that light pollution reduces the nocturnal pollinators' visits of flowers and thus impairs the plants‘ reproductive success. This altered insect behaviour also seems to affect the diurnal pollinator species, which are not able to compensate for the lost nocturnal 'work'.}

{As for humans, it has been established that the exposure to nocturnal light, especially at wavelengths around 440\,nm, reduces the concentration of melatonin in the blood, which acts as an anti-oxidant \citep{reiter2003melatonin}. Low melatonin levels have been linked to higher risk of cancer growth in humans \citep{blask2005melatonin}. Circadian disruption -- often triggered by artificial light -- also has other negative effects on human health (e.g.\ insomnia, depression and other psychiatric disorders, see \citet{wirz2009circadian}.)}

{Among animals, insects, migratory birds, hamsters and turtle hatchlings are some of the well documented cases where artificial light has particularly strong negative impacts \citep{gaston2015biological,BRV:BRV12036}}

{For studying the above mentioned hazardous effects of artificial light at night (ALAN), satellite images of the world (or individual countries) at night are not sufficient for various reasons, such as relatively poor spatial resolution. Where possible, they should be supplemented by ground-based measurements. Such ground-based measurements are the subject of the present paper, in which we investigate the NSB at 26 locations in Austria -- from heavily light-polluted to near-natural sites.}

{The structure and content of our paper is the following: In Sections\ 2-3, we describe our instrumental setup and the distribution of our measurement stations. Section 4 aims at the elaboration of suitable ways to graphically represent and analyse large sets of NSB measurements such as mean NSB histograms, annual ('hourglass') plots, density ('jellyfish') plots, derivation of hourly gradients of the NSB etc. In Sects.\ 5-6, we elaborate on the seasonal variations of the night sky brightness and its correlation with the aerosol content. Sect.\ 7 presents a brief analysis of the correlation between NSB and population, while Sect.\ 8 contains a discussion of long-term trends and the difficulties that we face when evaluating them. In the Conclusions, finally, we summarize our results and propose the establishment of a 'dark sky reserve' in Upper Austria.}


\section{Instruments and setup of our measurements}

{As mentioned in the previous section, the phenomenon of light pollution with all its consequences calls for monitoring programmes. These should ideally include NSB measurements on a global scale -- comparable to the global temperature measurements -- and with several techniques such as \textit{spectroscopy} of the night sky to track changes in the colour of skyglow, \textit{two-dimensional mapping} of the night sky from selected places with calibrated all-sky cameras in order to identify individual (growing or decreasing) 'light-domes' and, last but not least, one-dimensional, but \textit{time-resolved recordings of the zenithal NSB} from a grid of locations under all kinds of weather conditions and with automated data storage for 365 nights every year. While the methodologies of the different NSB measurement approaches has been explained in a recent review \citep{hanel2017measuring}, we based our present analysis solely on of these, i.e.\ zenithal NSB measurements performed with 'Sky Quality Meters' (SQMs) from the Canadian company Unihedron.}

\begin{figure}[h]
\centering
\includegraphics[width=1.0\textwidth]{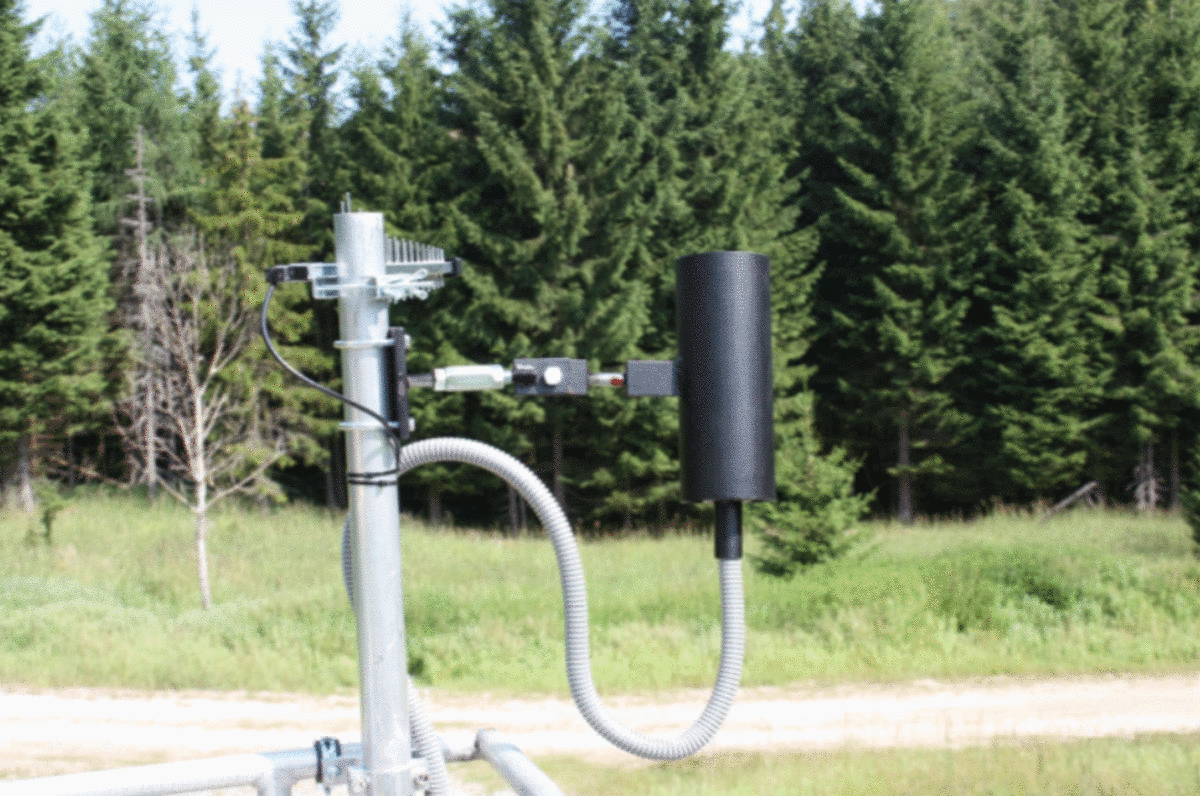}
\caption{Example for the setup of the Upper Austrian NSB measurement net at Z\"obelboden ({ZOE}). The black cylinder is the waterproof SQM housing.}
\label{setup}
\end{figure}

Similar networks for regular time-resolved NSB measurements have been established in other countries within the past few years as well. Examples include the stations established in Galicia \citep{bara2016anthropogenic}, in Catalonia \citep{phdthesis}, in Hong Kong \citep{Pun2014} as well as in the Netherlands \citep{denouter2015}. We chose this simple, but efficient measurement device for our measurement network in Eastern Austria. For a discussion of the advantages and shortcomings of this one-dimensional measurement approach, we refer again to the comprehensive review by \citet{hanel2017measuring}. All our SQMs are mounted in weatherproof housings (see Fig.\ \ref{setup}). {The weatherproof housings of the 23 SQMs in Upper Austria consist of black coated iron cylinders instead of the standard plastic case. To avoid pollution by bird dung, they have several sharp stings at the top.} We installed exclusively SQM-LE models, equipped with ethernet connections. Since these instruments have become well-known tools for the task of regular NSB measurements under both clear and cloudy (or even rainy) conditions, we abstain from a detailed technical description here (see \citet{cinzano2005night}).

Different sampling rates have been chosen for the individual SQMs: 7 seconds for the measurements at the Vienna University Observatory ({IFA}) as well as for those at Mount Mittersch\"opfl ({FOA}) in lower Austria (the respective instruments are operational since 2012); 90 seconds for the measurements at the observatory Graz-Lustbühel ({GRA}); and 60 seconds for all measurements done at the 23 stations in Upper Austria.\footnote{For the an overview of all stations and for the meaning of the station codes, see Sect.\ {3}.} The reasons for the different sampling rates are related to the different ways of data transfer and data storage.

The recorded data for {IFA, FOA and GRA} can be downloaded from the website
\url{https://www.univie.ac.at/nightsky},
while the data sets for Upper Austria will be found at \url{www.land-oberoesterreich.gv.at/159659.htm}.

\section{Distribution and description of the measurement stations}

As mentioned above, the measurements analyzed in this paper have been carried out at Vienna, Mount Mittersch\"opfl, Graz and at 23 locations in the county of Upper Austria. Some results of our measurements at the first two stations have been published by \citet{puschnig2014vienna}. The SQM station at Lustb\"uhel Observatory -- located at the outskirts of Graz -- was established in early 2014 and has been recording the NSB ever since. The extensive SQM network in Upper Austria has been set up with considerable efforts, starting in 2014, by the responsible county authority for environmental conservation, which is the reason for its large extent (23 stations, among them very remote ones as well as 13 urban ones). More than 10 million NSB measurements have been carried out in Upper Austria alone during 2015 and 2016.

\begin{figure}[]
\centering
\includegraphics[width=1.0\textwidth]{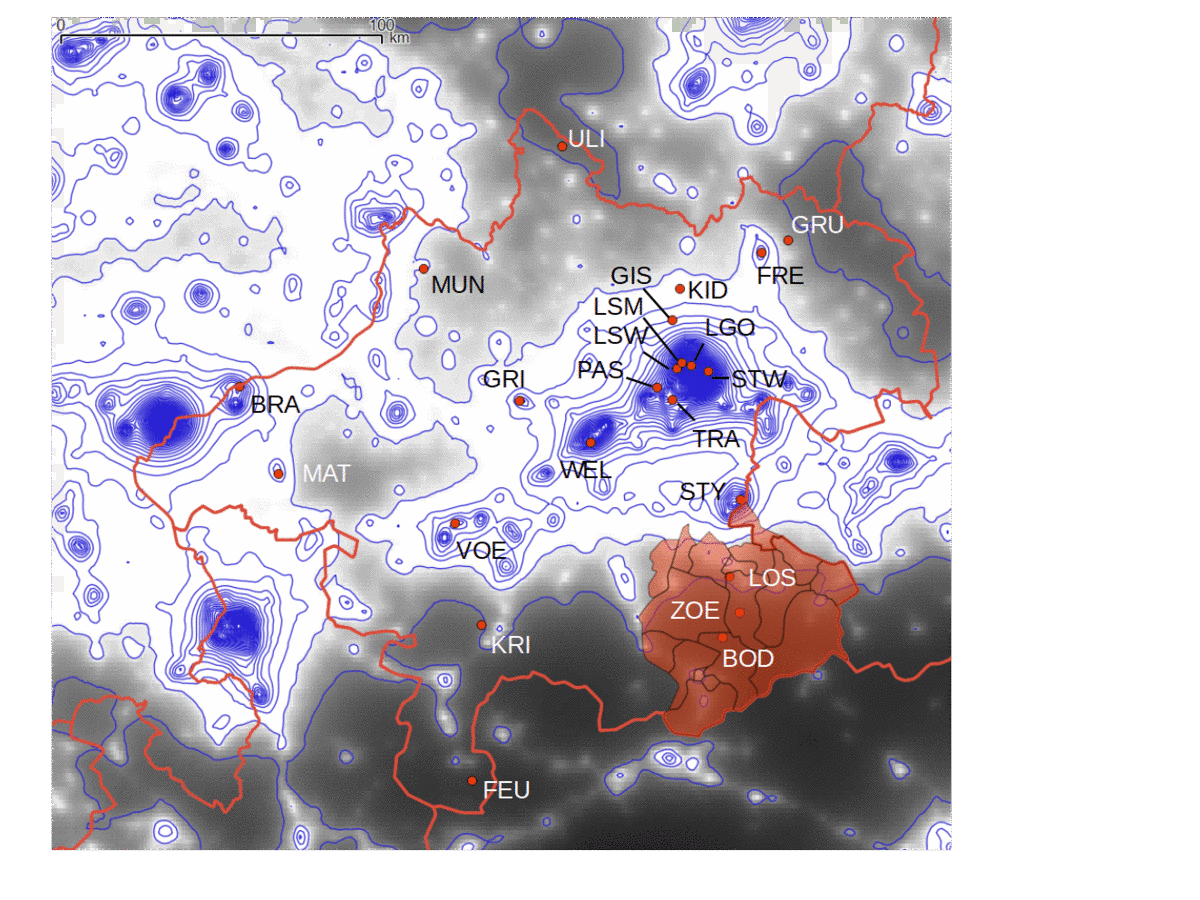}
\caption{Geographical distribution of the SQM stations in the Upper Austrian monitoring network. The red dots indicate their respective locations. The background map is based on the New World Atlas of Light Pollution by \citet{falchi2016,suppfalchi2016}. The stations Mittersch\"opfl {FOA}, Vienna {IFA} and Graz {GRA} are not shown in this image. The blue contour lines refer to the calculated artificial sky radiance at the zenith and have spacings of 46 $\mu$cd/m$^{2}$, starting at a baseline of 215 $\mu$cd/m$^{2}$, corresponding to 21.75 mag / arcsec$^{2}$, which is the requirement for an IDA ``gold'' tier for dark sky preserves. {For the meaning of the shaded area towards the lower right of the map, see end of Sect.\ 9}.}
\label{geodistribution}
\end{figure}

In the following, we give a description of each of the 26 stations. For an overview of coordinates and dates of establishment, see Tab.\ \ref{stations} and Fig.\ \ref{geodistribution}.

\newpage

\begin{itemize}

\item{BOD BODINGGRABEN}\\
This rural station is situated close to the border of the national park Kalkalpen. The location is in a narrow valley close to the rivulet 'Krumme Steyrling'. While the first by alphabet, this station was the last to become operable (end of June 2016). Altitude: 641\,{}m a.s.l.

\item{BRA BRAUNAU}\\
Urban station in the town of Braunau (population: 16.900). Dense residential area with the large river Inn passing in the north. Altitude: 351\,{}m a.s.l.

\item{FEU FEUERKOGEL}\\
Very remote rural station located at the top of the mountain Feuerkogel. Only scattered lodges and houses in the surroundings. Altitude: 1592\,{}m a.s.l.

\item{FOA MITTERSCHÖPFL}\\
Rural station located on top of Mount Mitterschöpfl, at the Leopold-Figl-Ob\-ser\-vatory of the University of Vienna, which hosts a 1.5 m and a 0.6 m telescope. It is surrounded by large forests which separates it from the nearest settlements. Altitude: 883\,{}m a.s.l.

\item{FRE FREISTADT}\\
Urban station in the city of Freistadt (population: 7.800) near the the city center and surrounded by dense residential areas. Altitude: 555\,{}m a.s.l.

\item{GIS GISELAWARTE}\\
Rural station at the rooftop of a look-out owned by the Austrian Alpine Association. The surroundings are uninhabited and dominated by dense forests. Altitude: 902\,{}m a.s.l.

\item{GRI GRIESKIRCHEN}\\
This urban station is situated on top of an office building close to the center of the town of Grieskirchen (population: 5.000) with dense residential areas nearby. Altitude: 336\,{}m a.s.l.

\item{GRA GRAZ}\\
Urban station located at the university observatory ``Lustb\"uhel'' at a distance of about 4,5 km from the city center of Graz (population: 284.000). It is surrounded by a park with modest amount of artificial lighting. Altitude: 484\,{}m a.s.l.

\item{GRU GRÜNBACH}\\
Rural station in north-eastern Upper Austria. South of it there is a loose settlement (population of the whole community: 1.900), while to the north farmland dominates. Some low traffic roads cross the area. Altitude: 918\,{}m a.s.l.

\item{IFA WIEN}\\
Urban station on the west terrace of the Vienna University Observatory, 3\,km from the city center. The observatory is surrounded by a forest, more than 5 hectares in size, with only very few artificial lights. The surroundings can be characterized as a residential area with moderate lighting levels. For technical reasons, data recording had to be paused from 3rd May 2015 to 21st April 2016. Altitude: 240\,{}m a.s.l.

\item{KID KIRCHSCHLAG-DAVIDSCHLAG}\\
Rural station, 30 km north of Linz, at the private astronomical observatory Davidschlag. The surroundings are dominated by agricultural fields and woods.
Apart from some farms in the environments, there is only little local influence of artificial light on the night sky brightness. Altitude: 813\,{}m a.s.l.

\item{KRI KRIPPENSTEIN}\\
Very remote rural station at the roof of the mountain station of a cable car. Alpine area without any permanent settlements. Part of the Dachstein plateau. In 2016, the data from KRI have a gap around midnight between July and December. Altitude of the station: 2067\,{}m a.s.l.

\item{LGO LINZ GOETHESTRASSE}\\
Urban station at a distance of 1.7\,km from the central square of Linz, the capital of Upper Austria (population: 203.000). The SQM has been mounted on a high rooftop. The surroundings are dominated by dense residential as well as business areas, administrative buildings and major roads. Altitude: 259\,{}m a.s.l.

\item{LSM LINZ SCHLOSSMUSEUM}\\
Urban station in the city of Linz, located on top of a hill, surrounded by green areas and parks. The SQM has been mounted on the rooftop of the ``Schlossmuseum'', only 300 m away from the central square. Altitude: 287\,{}m a.s.l.

\item{LSW LINZ STERNWARTE}\\
Urban station at the outskirts of Linz, more remote than the other two stations, located at a public astronomical observatory (``Johannes-Kepler-Sternwarte''), close to  residential and green areas. The distance to the city center amounts to 2 km. Altitude: 341\,{}m a.s.l.

\item{LOS LOSENSTEIN-HOHE DIRN}\\
Rural station on a slope on Mount ``Hohe Dirn'', uninhabited with woods and agricultural areas and some alpine cabins in the surroundings. The location is 6 km north of the northern border of the National Park ``Kalkalpen''. Altitude: 982\,{}m a.s.l.
 
\item{MAT MATTIGHOFEN}\\
Urban station in the market town of Mattighofen (population: 6.200), on the rooftop of the ``Schulzentrum'', surrounded by lose residential areas and cultivated lands. Altitude: 454\,{}m a.s.l.

\item{MUN MÜNZKIRCHEN}\\
{Rural} station in the village of Münzkirchen (population: 2.600), close to lose residential areas and agricultural fields with some forestation.
Altitude: 486\,{}m a.s.l.

\item{PAS PASCHING}\\
Urban station on top of the city hall of Pasching (population: 7.500). To the east and to the south there are agricultural areas while in the other directions residential areas are prevailing.  Altitude: 295\,{}m a.s.l.

\item{STY STEYR}\\
Urban station in the city of Steyr (population: 38.000), to the west of a business park. Towards the north, residential areas dominate. Altitude: 307\,{}m a.s.l.

\item{STW STEYREGG-WEIH}\\
Urban station 7 km to the south-east of the center of Linz. The station is located on a slope with green areas and forest in the vicinity. Even though Steyregg's population is less than 5.000, high NSB brightness values are prevailing due to the proximity of Linz. Altitude: 331\,{}m a.s.l.

\item{TRA TRAUN}\\
Urban station in the city of Traun (population: 24.300). The local environment includes a residential area, high-traffic roads, sports grounds, parking spaces, but also a lot of trees to the north. Altitude: 269\,{}m a.s.l.

\item{ULI ULRICHSBERG-SCHÖNEBEN}\\
Rural station located close to the market town of Ulrichsberg (population: 2.800), surrounded by scattered houses and forests. Altitude: 935\,{}m a.s.l.

\item{VOE VÖCKLABRUCK}\\
Urban station on a rooftop in the town Vöcklabruck (population: 12.300). Residential areas and major roads in the surroundings. Altitude: 434\,{}m a.s.l.

\item{WEL WELS}\\
Urban station on the rooftop of the city hall of Wels (population: 60.700), located within a dense residential area {directly at} the city center. Altitude: 317\,{}m a.s.l.

\item{ZOE ZÖBELBODEN}\\
This rural station is situated in the national park Kalkalpen. The surroundings are characterized by a forest glade with hilly terrain. No major settlements in the surroundings.
Altitude: 899\,{}m a.s.l.

\end{itemize}

\begin{table}[h]
\centering
\tiny
\caption{Coordinates, elevations, three-letter-codes and other details of our NSB monitoring stations, listed by categories (urban-intermediate-rural) and by alphabetic order. {All stations are still collecting data, but the present analysis is limited to the end date 2016-12-31.}}
\label{stations}
\begin{tabular}{llllcll} \hline \hline
Code & Name                  & Latitude N      & Longitude E      & Elevation [m]    & operational since \\
     &                           &                 &                  & (m above sea level) & (YYYY-MM-DD)    \\ \hline
     &                           &                  &  \{urban\}                    &                 \\

GRA  & Graz-Lustb\"uhel          & N 47° 4' 2"   & E 15° 29' 37"    & 484                 & 2014-01-01      \\
IFA  & Wien                      & N 48° 13' 54" & E 16° 20' 3"   & 231                 & 2012-03-01      \\
LSM  & Linz-Schlossmuseum        & N 48° 18' 19" & E 14° 16' 58"  & 287                 & 2014-09-06      \\
LGO  & Linz-Goethestraße         & N 48° 18' 19" & E 14° 18' 30"  & 259                 & 2014-09-06      \\
LSW  & Linz-Sternwarte           & N 48° 17' 36" & E 14° 16' 6"   & 336                 & 2014-09-06      \\
STY  & Steyr                     & N 48° 2' 57"  & E 14° 26' 32"  & 307                 & 2014-09-06      \\
STW  & Steyregg-Weih             & N 48° 17' 19" & E 14° 21' 13"  & 331                 & 2014-09-06      \\
TRA  & Traun                     & N 48° 14' 8"  & E 14° 15' 11"  & 269                 & 2014-12-05      \\
WEL  & Wels-Rathaus              & N 48° 9' 23"  & E 14° 1' 29"   & 317                 & 2014-09-06      \\

\hline
    &                           &   & \{intermediate\} &                     &                 \\
BRA  & Braunau                   & N 48° 15' 40" & E 13° 2' 41"   & 351                 & 2014-09-06      \\
GRI  & Grieskirchen              & N 48° 14' 4"  & E 13° 49' 33"  & 336                 & 2015-05-21      \\
FRE  & Freistadt                 & N 48° 30' 33" & E 14° 30' 7"   & 512                 & 2015-05-21      \\
MAT  & Mattighofen               & N 48° 5' 50"  & E 13° 9' 6"    & 454                 & 2014-12-15      \\
PAS  & Pasching                  & N 48° 15' 31" & E 14° 12' 36"  & 292                 & 2014-12-15      \\
VOE  & Vöcklabruck               & N 48° 0' 21"  & E 13° 38' 43"  & 434                 & 2014-12-15      \\

\hline
     &                           &   & \{rural\} &                     &                 \\
BOD  & Nationalpark-Bodinggraben & N 47° 47' 31" & E 14° 23' 38"  & 641 & 2016-06-29      \\
FEU  & Feuerkogel                & N 47° 48' 57" & E 13° 43' 15"  & 1628                & 2015-03-25      \\
FOA  & Mitterschöpfl             & N 48° 5' 3"   & E 15° 55' 24"  & 880 & 2012-11-16      \\
GIS  & Giselawarte               & N 48° 23' 3"  & E 14° 15' 11"  & 902                 & 2014-09-01      \\
GRU  & Grünbach                  & N 48° 31' 50" & E 14° 34' 30"  & 918 & 2014-09-06      \\
KID  & Kirchschlag-Davidschlag   & N 48° 26' 31" & E 14° 16' 26"  & 813  & 2014-12-15      \\
KRI  & Krippenstein              & N 47° 31' 23" & E 13° 41' 36"  & 2067 & 2015-11-04      \\
LOS  & Losenstein-Hohe Dirn      & N 47° 54' 22" & E 14° 24' 40"  & 982 & 2015-05-21      \\
MUN  & M\"unzkirchen             & N 48° 28' 45" & E 13° 33' 29"  & 486  & 2014-09-06 \\ 
ULI  & Ulrichsberg-Sch\"oneben     & N 48° 42' 20" & E 13° 56' 44" & 935                 & 2015-05-21      \\
ZOE  & Nationalpark-Z\"oblboden    & N 47° 50' 18" & E 14° 26' 28"  & 899                 & 2014-09-06      \\\hline
\end{tabular}
\end{table}


\section{Quantitative indicators and graphical representations of light pollution, demonstrated for selected measurement sites}

In this section, we present different ways of analyzing and representing our NSB measurements, with a focus on nocturnal averages (in the following: $<$NSB$>$) {and, geographically, with a focus on the data from Upper Austria}. The $<$NSB$>$ values are defined as arithmetic means of the individual NSB values for each night, {calculated for solar elevations below -15 degrees. Note that the natural sky brightness hardly changes below this solar elevation.}

One might object that NSB values, expressed in magnitudes and hence in logarithmic units, are not suitable for calculating arithmetic mean values. However, the $<$NSB$>$ values calculated in this way do have a 'correct' meaning, since arithmetic averages of logarithmic values are equivalent to geometric averages of the non-logarithmic values. In both cases, smaller numbers (= brighter skies) are more strongly weighted than larger numbers, while this is of course not the case for arithmetic averages of non-logarithmic values. 
This does not invalidate the applied procedure of calculating $<$NSB$>$, but needs to be kept in mind in the following.

\subsection{Mean NSB histograms}

A very appropriate way to characterize the darkness or light pollution at a given site is its $<$NSB$>$ histogram plot. In this representation, the $<$NSB$>$ values are split into equal-sized bins (bin size = 0.2 mag in our case), and for each bin the number of points from the data set that fall into each bin is shown on the y axis. The total frequency is plotted on the left y axis label. In addition, we calculated the normalized frequency of $<$NSB$>$ values up to each magnitude bin and represented the resulting curve in dark green. In this additional representation, the final value of the frequency, at the cutoff value of 22.6 mag$_{SQM}$/arcsec$^{2}$, always reaches 1.0.

Since all our measurement stations reliably collecting data for the whole year 2016\footnote{Except for the station BOD, which is collecting data only since end of June 2016.}, we created the $<$NSB$>$ histograms for this year, namely for each station. {The results, arranged by increasing average NSB, are shown in the Appendix, Fig.\ \ref{hist1}-\ref{hist2}.}

\begin{figure}[]
\centering
\includegraphics[width=1.0\textwidth]{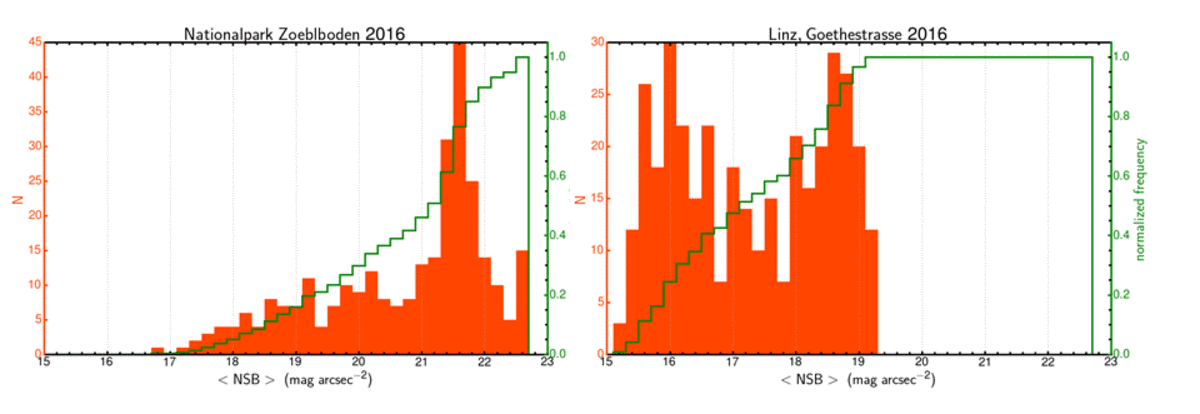}
\caption{Mean NSB histograms for the stations Z\"oblboden ({ZOE}) and Linz-Goethestra{\ss}e ({LGO}) based on measurements from 2016. The left histogram exhibits a single outstanding maximum, namely at 21.6 mag$_{SQM}$/arcsec$^{2}$, corresponding to the distinguished condition of clear nights with very little, if any, contribution of scattered moonlight to the sky brightness. In the right histogram, two peaks are clearly visible, one at 16.0 and one at 18.6\,mag$_{SQM}$/arcsec$^{2}$, corresponding to the mean sky brightness in overcast nights and in clear, moonless nights.}
\label{histograms}
\end{figure}

In the following, we shall discuss two selected histograms, both contained in one figure
and representing interesting extreme cases.
The left half of Fig.\ \ref{histograms} contains a histogram of $<$NSB$>$ values derived at the rural station {ZOE}. Due to the very small amount of light pollution there, the dominant frequency in this histogram is at 21.6 mag$_{SQM}$/arcsec$^{2}$. We assume that this corresponds to the typical zenithal sky brightness at this site under clear conditions with only a small influence of moonlight. The histogram shows that darker values of $<$NSB$>$ also occur at {ZOE}, which either correspond to no moonlight at all or -- especially beyond 22\,mag$_{SQM}$/arcsec$^{2}$ -- to overcast nights. Note that this is one of the relatively rare documented cases where overcast nights may get \textit{darker} than clear nights.
{Similar histograms have been found for the rural for the stations BOD, LOS, KRI, FEU, GRU, KID and ULI (see Appendix, Fig.\ A1).}

The right half of Fig.\ \ref{histograms}, in contrast, represents a totally different distribution of $<$NSB$>$ values. It was derived from the measurements at the urban station LGO. At this station in Linz -- and at all other urban sites, see again Fig.\ \ref{hist2} --, clear nights are firstly much brighter than in rural regions, in this case corresponding to the right frequency maximum, located at 18.6\,mag$_{SQM}$/arcsec$^{2}$. This means about 11 times brighter skies than at {ZOE} under comparable conditions. Secondly, cloudy and overcast nights are even much brighter still, {as has been shown, inter alia, by \citet{kyba2012} and by \citet{so2014}.}
We take this additional night sky brightening by clouds as the (main) explanation of the left frequency maximum in the right half of Fig.\ \ref{histograms}, located at 16.0\,mag$_{SQM}$/arcsec$^{2}$ -- which is 250\,times the natural zenithal NSB
(assumed here as 22.0\,mag/arcsec$^{2}$).


\subsection{Regularity of the circalunar $<$NSB$>$-variation}

Another way of analyzing and representing the amount of light pollution at a site is by generating plots of the circalunar variation of the mean NSB. The more pristine the sky a site, the stronger is the correlation between the $<$NSB$>$ and the lunar phase. In the diagrams that we generated -- again focusing on the year 2016 --, full moon nights are denoted by solid (light blue) vertical lines, while the times of new moon are indicated by dashed (black) vertical lines.

\begin{figure}[]
\centering
\includegraphics[width=1.0\textwidth]{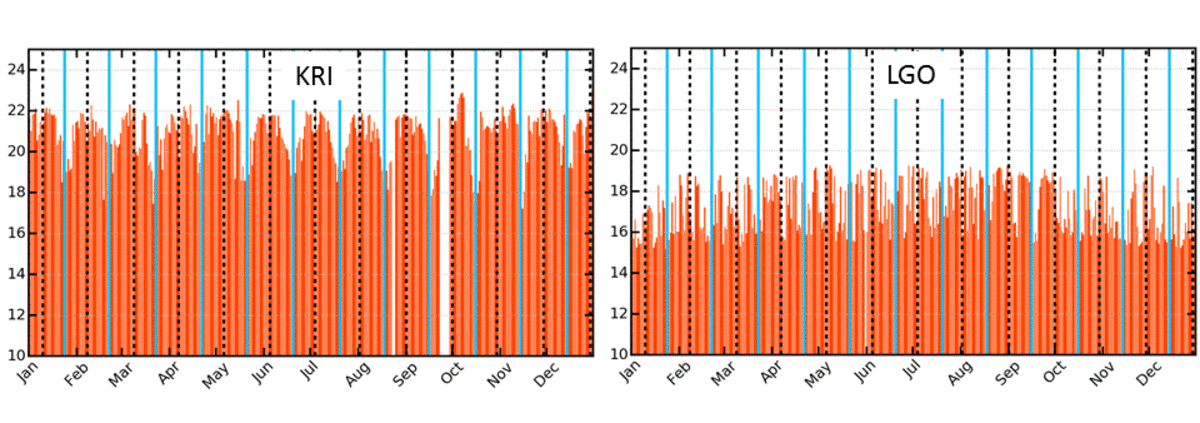}
\caption{Variation of the $<$NSB$>$ with the lunar cycle in 2016 at the stations KRI and LGO. Dotted vertical lines denote the times of New Moon. Full lines denote Full Moons.}
\label{circalunar2016}
\end{figure}

Figure \ref{circalunar2016} contains {two} circalunar plots in one, this time using the examples of {KRI and LGO}. The darkness of each night is characterized by the height of the narrow many red (in the printed version: grey) vertical lines extending up to different values. For the right hand side case, {Linz (LGO)}, no single $<$NSB$>$ line extends up to more than 19.2 mag$_{SQM}$/arcsec$^{2}$, while cloudy nights, especially in winter, can reach as bright mean zenithal sky brightness values as 15.5 mag$_{SQM}$/arcsec$^{2}$. At the same time, the regularity of the $<$NSB$>$ is very poor. The pattern appears almost stochastic, since the main ``driving force'' of sky brightness is the cloudiness, which follows no strict periodicity.
The opposite is true for Krippenstein ({KRI}), our most remote station, located at a height of more than 2.000\,{}m a.s.l.\ on the Dachstein plateau. Here, the circalunar periodicity is very strong, and
$<$NSB$>$ values beyond 22\,mag$_{SQM}$/arcsec$^{2}$ occur, not only close to New Moon, but with arbitrary lunar phases during overcast nights.
Clear full moon nights, especially in winter, can become as bright as 17.3 mag$_{SQM}$/arcsec$^{2}$ in $<$NSB$>$, which implies a total amplitude close to 5 magnitudes -- almost a factor of 100 in luminance. 

 {To give a more complete picture of the circalunar $<$NSB$>$ variation, we generated the corresponding graphs for all 23 Upper Austrian stations and collected them in the Appendix, Figs.\ \ref{circalunar1}--\ref{circalunar2}. As can be seen there, high  $<$NSB$>$ amplitudes and strong correlations with the lunar phases are found for all remaining rural stations in Upper Austria: BOD, ZOE, LOS, KRI, FEU, GRU, KID, ULI, GIS and MUN. For the 'intermediate' (MAT-PAS) and 'urban' stations (STY-LGO), successive brightening and in parallel a transition to stochastic $<$NSB$>$ patterns will be noted.}
 
 {For a quantitative analysis of the $<$NSB$>$ periodicity, we did a series of Fourier transforms that yielded very clear and promising results. Their presentation would, however, lead too far astray here, and will be the subject o a forthcoming paper (Puschnig et al., in preparation).}

\subsection{'Hourglass' plots}

While the previous figures and sections were based on analyses of \textit{mean} NSB values, there are other ways of representing the temporal (esp.\ circalunar) variation of the night sky brightness, without the need of averaging the values for each night.

One particularly instructive kind of diagram is the 'hourglass' plot, which has been introduced in previous papers, e.g.\ \citet{bara2016anthropogenic}, \citet{ribas2016}.

In this case, as for the previous circalunar periodicity diagrams, the x axis is a time axis, containing the months of one full year. However, the y axis is a time axis as well, but covering the much shorter timescale of the hours (and fractions of hours) of the individual nights. A colour or grey scale is used to denote the measured NSB at each time of the night and of the year.

Again, the circalunar NSB periodicity or a lack of periodicity can be well recognized in the resulting plots. But in addition, other features emerge, e.g.\ the natural variation of the night lengths, which creates the 'hourglass' shape and is obviously more pronounced for higher latitudes with their extremely short summer nights. 

\begin{figure}[ht]
\centering
\includegraphics[width=1.0\textwidth]{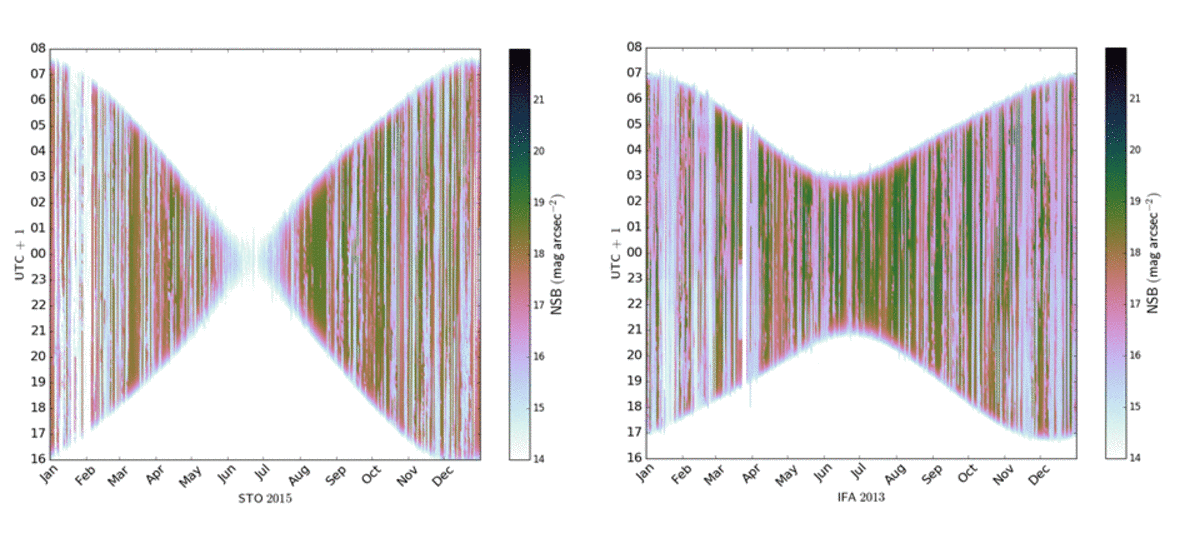}
\caption{Left: Illustration of an extremely strong annual variation, but strongly distorted circalunar NSB pattern by an hourglass plot for the data from the Stockholm University Observatory from 2015. Right: Hourglass plot for the data from the Vienna University Observatory (IFA) from 2013.}
\label{STOVIE}
\end{figure}

To illustrate the latter point, we start here with a plot from a place far beyond our measurement network, namely Stockholm (59$^{\circ}$ 20' northern latitude or 7 degrees south of the northern arctic circle). As the left half of Fig.\ \ref{STOVIE} demonstrates, only a maximum darkness of $\sim$14\,mag$_{SQM}$/arcsec$^2$ is reached at this place during the summer solstice. The measurements at the University Observatory of Stockholm have been collected by one of us (JP), who also wrote the Python scripts to generate all the remaining plots shown in this section and in the Appendix. Note that a single hourglass plot may contain up to several 100.000 individual data points, depending on the sampling rate.
For comparison, the right half of Fig.\ \ref{STOVIE} shows the hourglass diagram of the NSB data measured at IFA (Vienna University Observatory) in 2013. Here, summer nights are obviously not as short as at Stockholm, which is why the central (summer) part of the hourglass is definitely broader in this case. Both cities, however, have similar NSB distributions with respect to the absence of the circalunar pattern.

\begin{subfigures}
\begin{figure}\centering
  \includegraphics[width=1.0\textwidth]{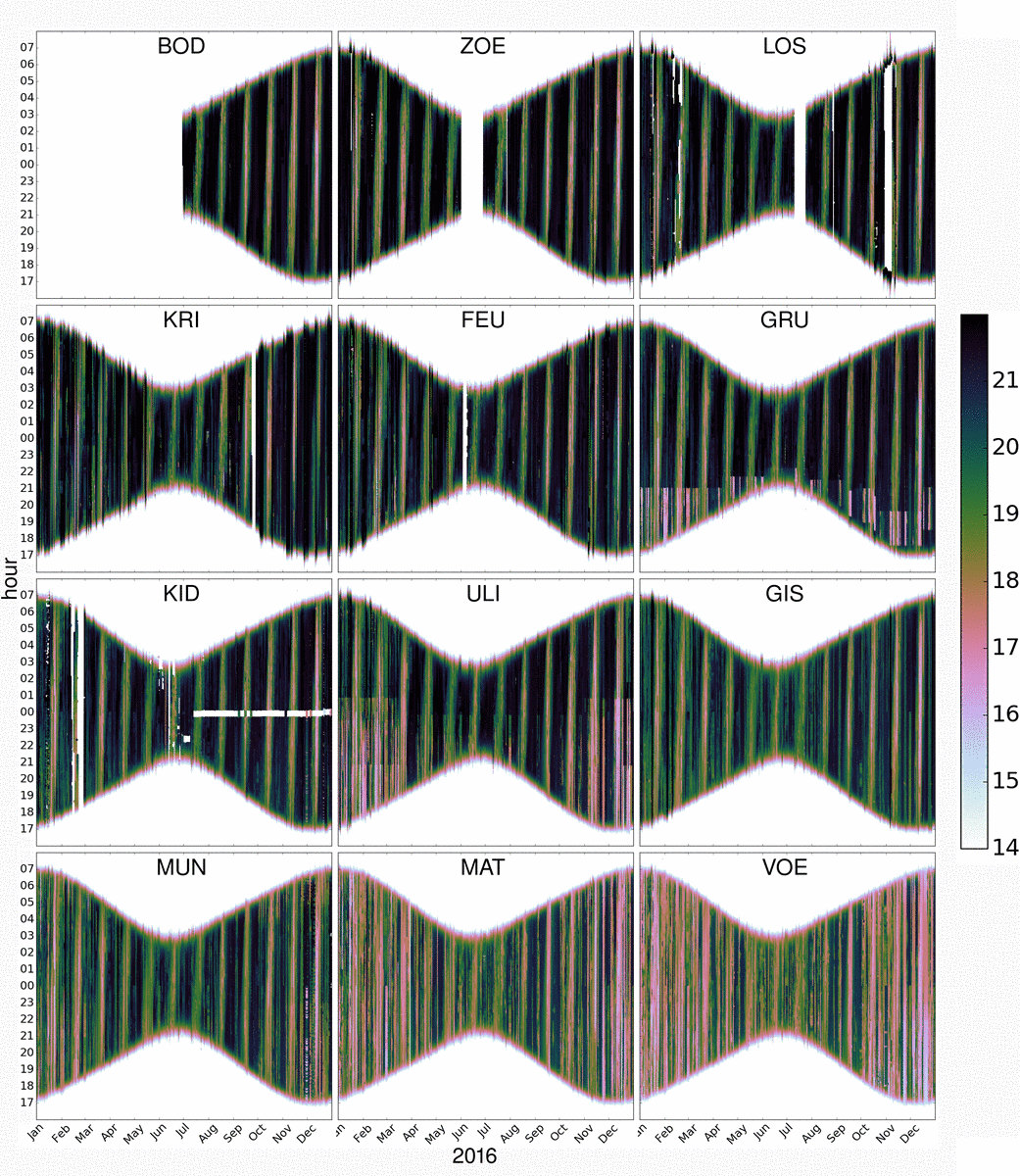}
  \caption{\label{hourglass2016_1}Hourglass plots of the NSB in 2016 at the 23 measurement sites in Upper Austria (part 1: BOD-VOE). Measurements at BOD started only in the second half of 2016.} 
\end{figure}
\begin{figure}\centering
  \includegraphics[width=1.0\textwidth]{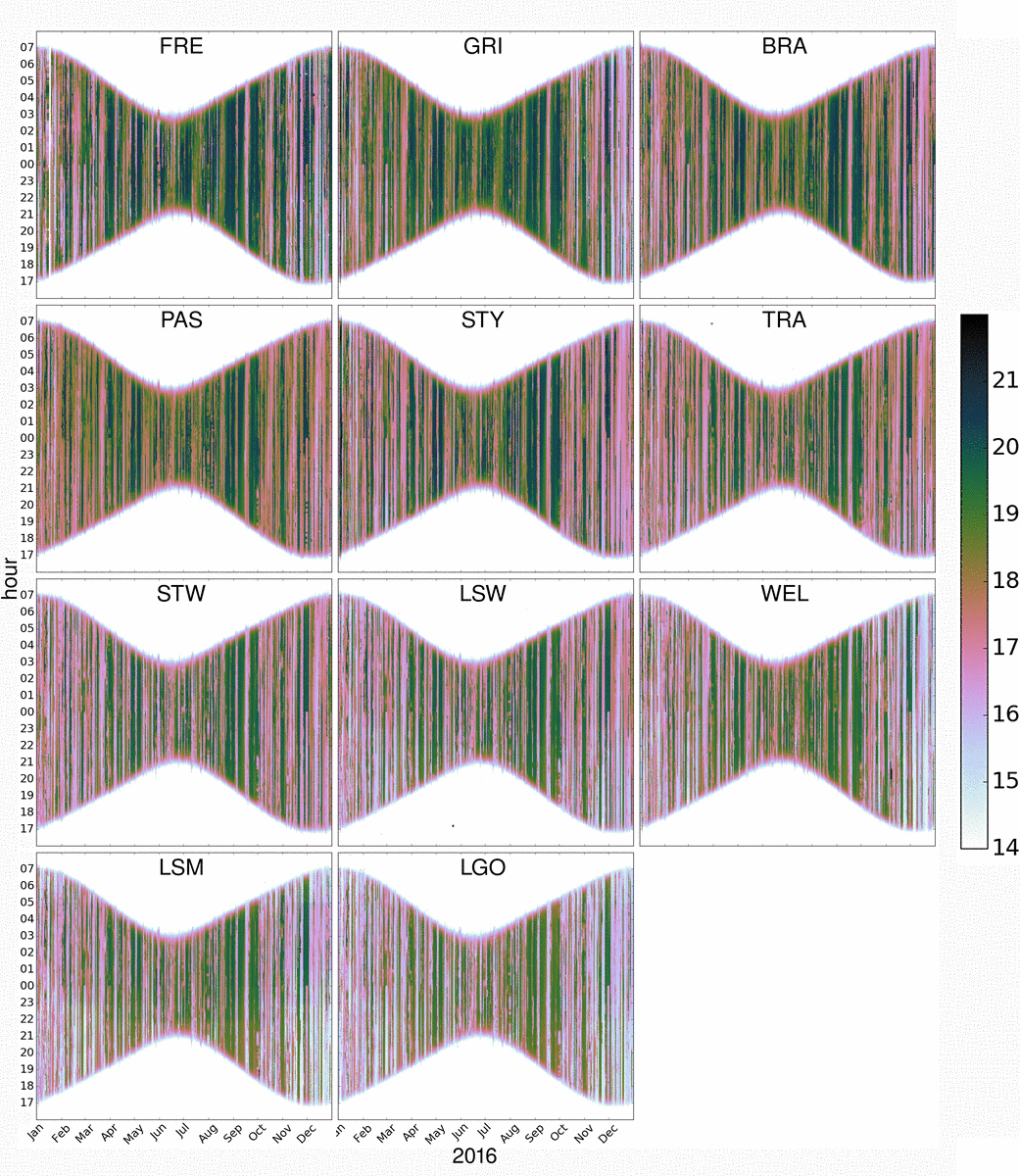}
  \caption{\label{hourglass2016_2}Hourglass plots of the NSB in 2016 at the 23 measurement sites in Upper Austria 
  (part 2: FRE-LGO). Note the progressive degradation of the circalunar periodicity at these mostly urban sites.} 
\end{figure}
\end{subfigures}

{For Upper Austria, we show hourglass plots for all 23 measuring stations in this county for the year 2016 in Fig.\ 6. The individual plots are ordered starting with the darkest sites and progressing toward the most light-polluted stations (i.e.\ in the same way as in Tab.\ \ref{nsbmax-t}). The first three plots (for BOD, ZOE and LOS) unfortunately show large gaps in the data acquisition, which is partially due to the remote position of the corresponding stations. For these and the remaining rural stations (KRI-MUN), the circalunar periodicity pattern is very clearly seen. For the 'intermediate' stations according to Tab.\ 1 (MAT, VOE, FRE, GRI, BRA and PAS), at least for the summer months, a weak co-variation of the NSB with the lunar phases can be recognized. At the same time, the annual plots become more and more 'noise'-dominated, i.e.\ meteorological effects start to take over gradually. Finally, in the hourglass plots for the 'urban' stations in Upper Austria (STY, TRA, STW, LSW, WEL, LSM and LGO), the circalunar pattern vanishes entirely. In addition, a trend toward brighter night skies during the winter months can be seen in all plots from MAT to LGO (i.e.\ for all stations except the rural ones). Some stations show significantly brighter NSB values until a certain time of the night, e.g.\ 9 PM, 10 PM, 12 PM or 1 AM. This is the case for the stations GRU and ULI. We assume that local stray light from a source that is being turned off at some point is responsible for these artifacts.
For the urban stations in large agglomerations, a gradual darkening of the night sky can be spotted in the hourglass plots. We shall come back to this nocturnal later (e.g.\ Tab. \ref{nsbmax-t}). Finally, we note that one of the stations (KID) shows a 'horizontal' data gap around midnight between July and December 2016.}


\subsection{Density plots ('jellyfish' plots)}

The usefulness of density plots for the characterization of light-polluted sites has already been demonstrated in various papers, e.g.\ \citet{bara2016anthropogenic}, \citet{puschnig2014vienna}, \citet{puschnig2014potsdam}, \citet{so2014}, \citet{ribas2016}.

\begin{figure}[]
\centering
\includegraphics[width=1.0\textwidth]{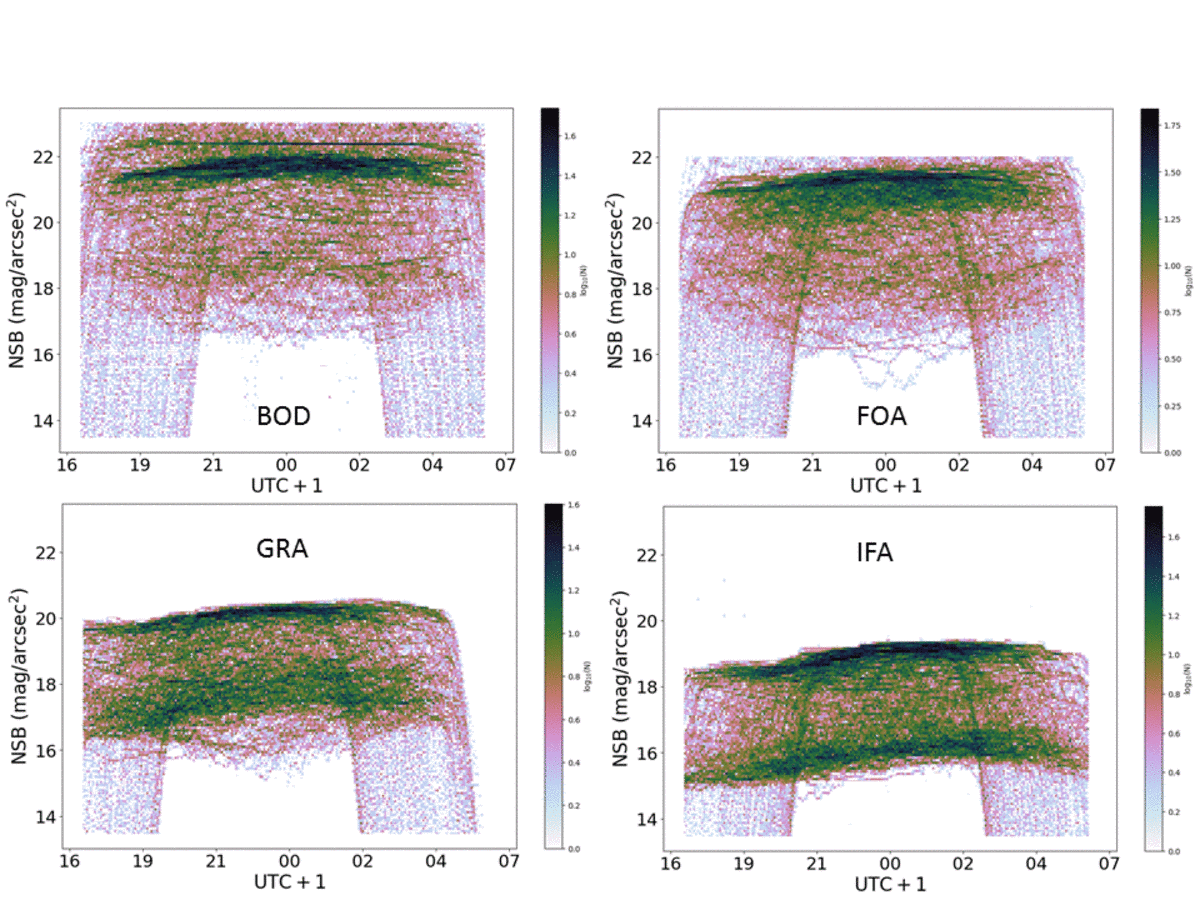}
\caption{Jellyfish diagrams of the NSB recorded in 2016 at the stations BOD, FOA (both rural), GRA and IFA (both urban). Note the positive nocturnal gradients of the NSB at the urban stations.}
\label{jellyfish}
\end{figure}

For convenience, we shall call these diagrams 'jellyfish' plots in the following -- due to the corresponding overall shapes of the emerging point clouds (see Fig.\ \ref{jellyfish}). Jellyfish plots for urban, light-polluted sites show two cluster regions, which have little to do with the lunar phases, but correspond to clear nights with moderate skyglow on the one hand and overcast nights with strongly enhanced scattering of the city lights on the other hand. In the lower half of Fig.\ \ref{jellyfish}, the two cluster regions are quite bluntly seen. In the case of Graz-Lustb\"uhel (GRA), the vertical distance between the two cluster regions -- the lower (overcast) and the higher (clear) region -- amounts to about 2.7 mag$_{SQM}$/arcsec$^{2}$.

This value closely approaches the one previously found by \citet{puschnig2014potsdam} for Potsdam-Babelsberg and is also similar as for the site Vienna University Observatory (IFA, lower right plot).
This means that in Graz, Vienna and Potsdam the overcast sky is, on average, 13 times brighter than the clear sky. 

We may introduce the term 'spread' for the vertical distance between the middle line through the cluster region of overcast nights and the much more compressed cluster region of clear nights. This spread can also be recognized in many of the histograms shown in the Appendix, but there it shows up as the \textit{horizontal} distance between the two strongly dominant peaks of the density distribution.

While a large 'spread', \textit{up to almost 3 magnitudes}, occurs in large cities, it practically vanishes at stations with very small zenithal luminance such as BOD and FOA in the upper half of Fig.\ \ref{jellyfish}. In other words, there is no significant enhanced backscattering of artifical light close to the zenith at such rural places, at least not to the point as to generate an overcast NSB cluster \textit{below} (= to the \textit{brighter side} of) the clear sky NSB cluster, but only the latter emerges in the respective jellyfish plot. However, scattered NSB values \textit{above} (=to the \textit{darker side} of) the clear sky NSB cluster show up in the diagrams for such hardly light-polluted regions. The most likely explanation for most of these scattered dark NSB values -- reaching values beyond 22\,mag$_{SQM}$/arcsec$^{2}$ as in some of the histograms -- is \textit{darkening of the night sky by clouds}, a phenomenon that is rarely seen in Europe nowadays but that has been reported by other authors as well -- see also \citet{ribas2016}, \citet{jechow2016evaluating}.

Weak clustering of NSB values in the range of about 18.5\,mag$_{SQM}$/arcsec$^{2}$, which is also seen at the national park site BOD (and at comparably dark sites) probably corresponds to moonlit skies at different lunar phases.

Another evident feature in the jellyfish plots for urban regions is the increase (=darkening) in the NSB between the evening and morning hours. This creates a significant left-half vs.\ right-half-asymmetry in the cumulative NSB diagrams. For natural conditions, the diagrams should be very close to axially symmetrical with respect a vertical line at local midnight. Such a symmetry is almost reached for the station {BOD} (upper left diagram in Fig.\ \ref{jellyfish}).

A complete set of jellyfish plots for all 23 NSB measurement stations in Upper Austria will be found in the Appendix, Figs.\ \ref{jelly1}-\ref{jelly2}.


\subsection{Averaged jellyfish plots and nocturnal development of the NSB}

In order to study the mean magnitude and time-evolution of the NSB under \textit{all} meteorological conditions for a given site, it does make sense to calculate the yearly average NSB for those times around midnight ($\pm$ 2-3 hours) where darkness is reached throughout the year. The result corresponds to a mean, representative scotograph from all individual scotographs characterizing the NSB(t) for each night. One may equally conceive of the result as a density-weighted 'midline' through the above-mentioned jellyfish diagrams -- therefore we may use the term 'averaged jellyfish plots' in this context. Qualitatively similar plots -- but with a focus on moonlight-free nigths -- have previously been produced and applied by \citet{so2014} (e.g.\ Fig.\ 3.22) in order to characterize the light pollution at a number of sites within Hong Kong.

Two useful quantities may be derived from averaged jellyfish plots: 

\begin{itemize}
    \item{'NSB$_{avg, max}$', the location of the upper horizontal envelope of each averaged jellyfish graph}
    \item{'t-grad', the hourly gradient of NSB(t), derived by drawing a line connecting the values at 10\,PM and 2\,AM}
\end{itemize}

The first parameter characterizes the maximum darkness that is reached at a given site \textit{on average}, i.e.\ during all kinds of conditions (moonlit, moon-free, clear, overcast) in the course of the night. The second parameter -- which includes some arbitrariness in its definition (esp.\ the linear approximation) -- is a rough measure of the extent to which zenithal skyglow decreases per hour, again at a given site and again for all covered meteorological conditions.

\begin{figure}[]
\centering
\includegraphics[width=1.0\textwidth]{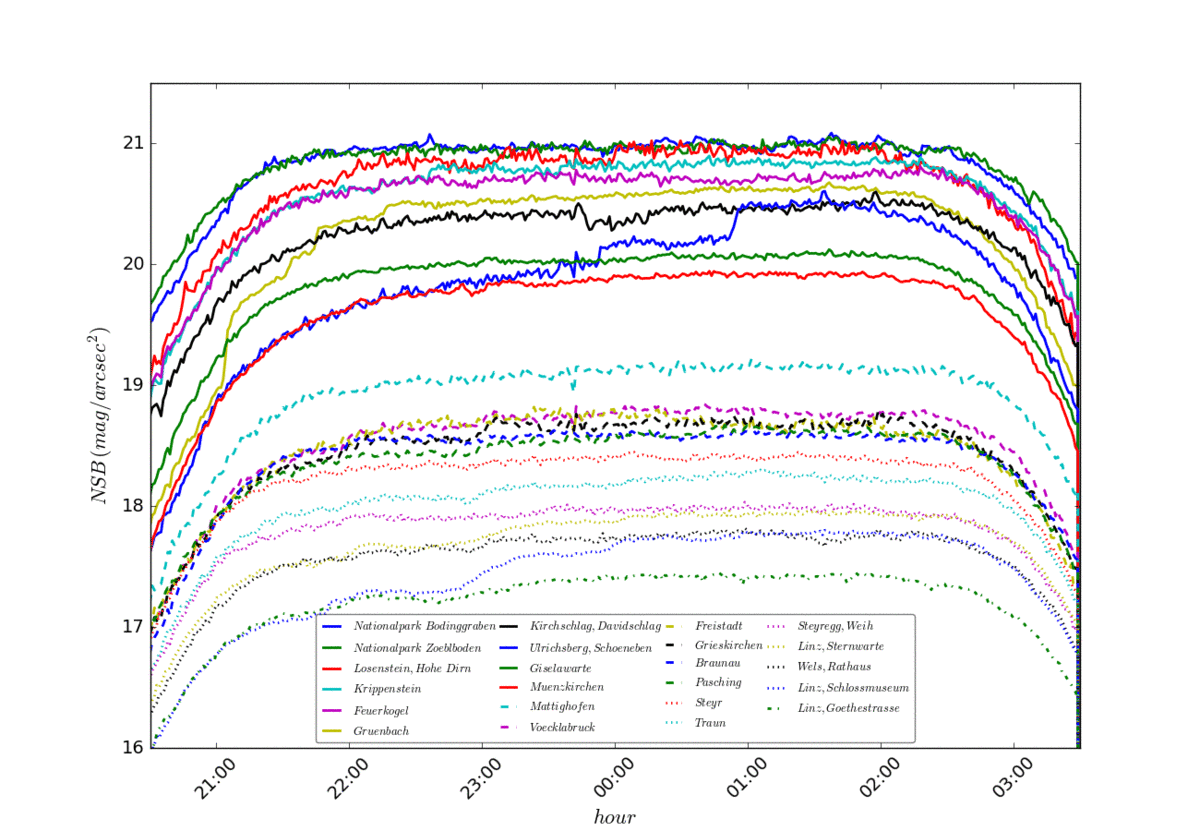}
\caption{Averaged jellyfish diagrams of the NSB, recorded in 2016 at all stations located in Upper Austria. Note the trend of decreasing nocturnal NSB-gradients ('t-grad') upon transition from urban to rural stations (i.e.\ from the bottom to the top).}
\label{avgscoto}
\end{figure}

Figure \ref{avgscoto} shows the averaged jellyfish plots for all our stations in Upper Austria (based on the measurements in 2016). {The most conspicuous feature of this diagram is the vertical distance of almost 1 magnitude between the rural stations (BOD down to MUN) and the intermediate as well as urban stations (MAT down to LGO).}

Table \ref{nsbmax-t} gives an overview of the corresponding 'NSB$_{avg, max}$' values (second column). We compare these values with the $<$NSB$>$ reached at the respective same stations during clear, moonless nights ($<$NSB$>$ (clear), third column). It should be noted that the NSB$_{avg, max}$ reach 21.1 mag$_{SQM}$/arcsec$^{2}$ at best, as they result from a massive averaging process that does \textit{not} only include clear and moonless nights, as mentioned above. By contrast, and fortunately, the $<$NSB$>$ (clear) values reach 21.8-21.9 mag$_{SQM}$/arcsec$^{2}$ for some of our sites. These sites could reach the 'gold tier' awarded by the International Dark Sky Association for very pristine Dark Sky Reserves. Note, furthermore, that the difference between NSB$_{avg, max}$ and $<$NSB$>$ (clear) is not the same for all stations, but slightly increases from 0.7 to 1.5 magnitudes as we proceed from rural to urban sites. This is because NSB$_{avg, max}$ gets more and more influenced by the bright clouded skies at urban sites, while this (frequently occurring) condition is not reflected by the $<$NSB$>$ (clear) values.
As for the last column of Tab.\ \ref{nsbmax-t}, it contains the `t-grad' values, the meaning of which has been explained above. Here again, a transition from small (even vanishing) values at extremely remote sites to larger nocturnal gradients in cities like Linz ({LGO, LSM, LSW}) and Wels ({WEL}) is noteworthy.

For Vienna ({IFA}), we find an average nocturnal increase in the zenithal sky brightness in the order of 0.1\,mag$_{SQM}$/arcsec$^{2}$, confirming an earlier result by \citet{puschnig2014vienna}.

\begin{table}[]
\centering
\small
\caption{Average darkness that is reached, at the darkest hour, under all conditions at the 23 measurement stations in Upper Austria (col.\ 2) versus darkest $<$NSB$>$ that is typically reached at the respective sites during \textit{clear, moonless} nights (col.\ 3; these values may vary by $\pm$ 0.2\,mag/arcsec$^{2}$ depending on the season, visibility of the Milky Way etc.). The last column contains the hourly increase rate of the NSB between 10 PM and 2 AM ('t-grad').}
\label{nsbmax-t}
\begin{tabular}{llll} \hline \hline
Site & NSB$_{avg, max}$ &  $<$NSB$>$ (clear) & t-grad  \\
& [mag/arcsec$^{2}$] & [mag/arcsec$^{2}$] & [mag/hour] \\
\hline
BOD & 21.1 & 21.8 & 0.0 \\
ZOE & 21.1 & 21.9 & 0.0 \\
LOS & 21.0 & 21.8 & 0.0 \\
KRI & 20.9 & 21.9 & 0.0 \\
FEU & 20.8 & 21.8 & 0.0 \\
GRU & 20.7 & 21.6 & 0.0 \\
KID  & 20.6 & 21.6 & 0.0 \\
ULI & 20.6 & 21.6 & 0.0 \\
GIS & 20.1 & 21.3 & 0.0 \\
MUN & 20.0 & 21.2 & 0.01 \\
MAT & 19.2 & 20.8 & 0.05 \\
VOE & 18.8 & 20.4 & 0.05 \\
FRE & 18.7 & 20.2 & 0.04 \\
GRI & 18.7 & 20.2 & 0.05 \\
BRA & 18.7 & 20.2 & 0.03 \\
PAS & 18.7 & 20.1 & 0.05 \\
STY & 18.5 & 20.3 & 0.04 \\
TRA & 18.3 & 19.8 & 0.06 \\
STW & 18.0 & 19.7 & 0.07 \\
LSW & 18.0 & 19.5 & 0.07 \\
WEL & 17.8 & 19.4 & 0.09 \\
LSM & 17.8 & 19.2 & 0.14 \\
LGO & 17.7 & 19.0 & 0.12 \\ 
\hline
\end{tabular}
\end{table}

In Tab.\ \ref{nsbmax-t}, we give an overview of the NSB$_{avg, {max}}$ and t-grad values, derived from the measurements performed in year 2016 at all the stations located in Upper Austria. Figure \ref{avgscoto} shows the corresponding averaged jellyfish diagrams.


\section{Seasonal variations of the mean NSB}

In addition to the nocturnal means of the NSB, we calculated monthly averages (= mean values of the nocturnal $<$NSB$>$ values), including all kind of weather conditions and all lunar phases. This made it possible to study seasonal variations of the night sky brightness.

It turned out that the amplitude of these variations is much larger at urban sites, i.e.\ typically 1.5 magnitudes, while it is small at rural sites (typically 0.5 magnitudes). In both cases, the darkest NSB values are reached in summer (between July and September).

\begin{figure}[]
\centering
\includegraphics[width=1.0\textwidth]{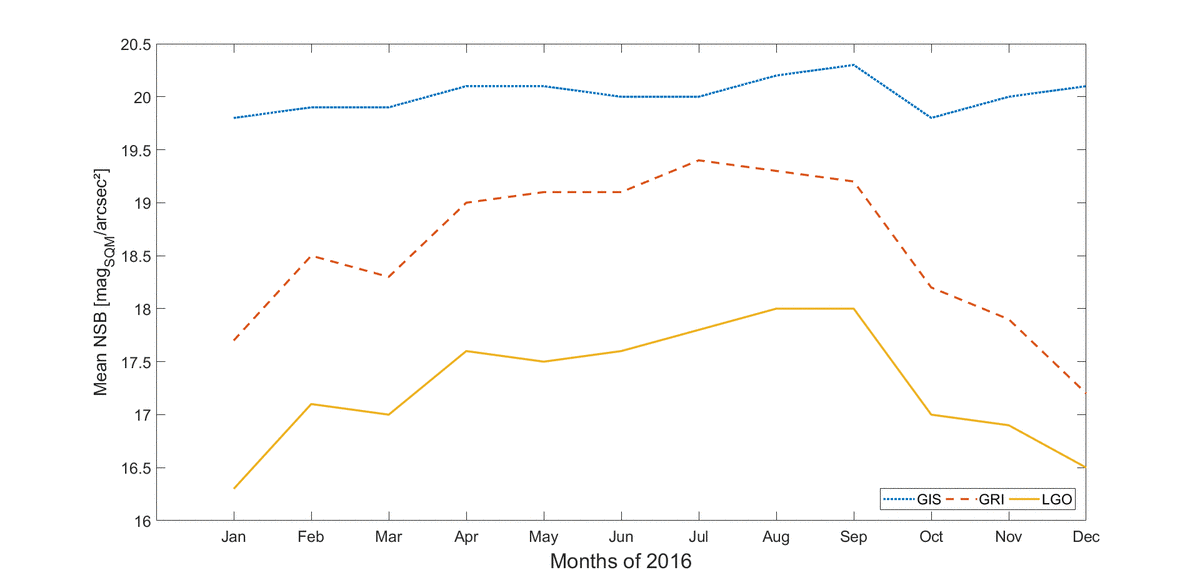}
\caption{Monthly averages of the NSB for three stations measured during the year 2016. The bottom two graphs are from densely populated areas (Linz-Goethestraße {LGO}, Grieskirchen {GRI}), while the graph at the top is from a rural station (Giselawarte {GIS}). While the two bottom graphs show pronounced seasonal variations, the rural station shows a rather moderate one.}
\label{3stations}
\end{figure}

In Fig.\ \ref{3stations}, the seasonal variation of the mean NSB at the three stations {LGO} (urban), {GRI} (intermediate) and {GIS} (rural) is shown and the above mentioned trend is clearly seen in two of the data sets. 

\begin{figure}[]
\centering
\includegraphics[width=1.0\textwidth]{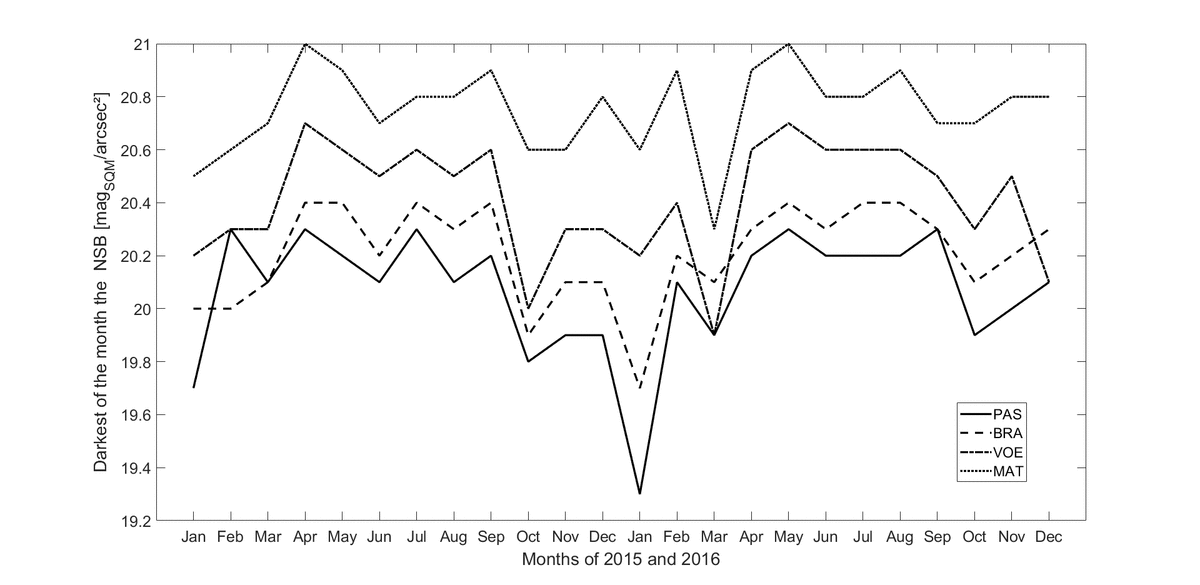}
\caption{Darkest values of the NSB per month for four stations (PAS, BRA, VOE and MAT) during the years 2015 and 2016. Note the similarity in the overall course of the seasonal NSB variations.}
\label{longdark}
\end{figure}

Comparing different years lead us to the conclusion that the seasonal trend in the NSB is roughly the same on longer timescales -- see Fig.\ \ref{longdark}. Here again, the skies over all four stations consistently show brightness records during the winter months December and January.

\begin{figure}[]
\centering
\includegraphics[width=1.0\textwidth]{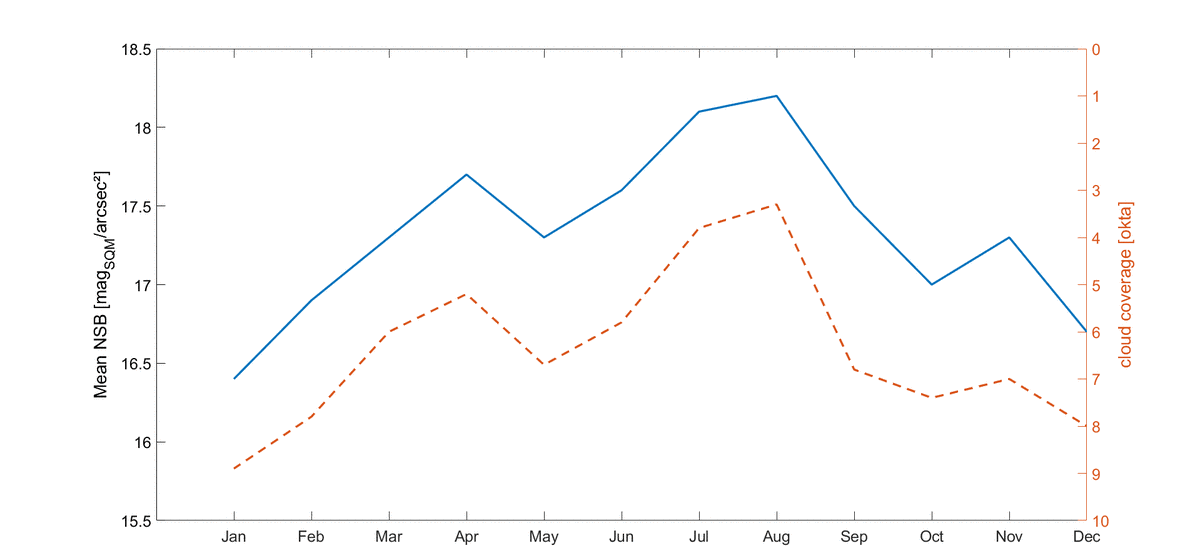}
\caption{Monthly averages of the NSB (solid line) versus cloud coverage at the station Linz-Goethestra{\ss}e ({LGO}) in 2015. The right y-axis is inverted since lower okta values correspond to darker skies in cities. The correlation between the two plotted parameters confirms that clouds have a dominant effect on the NSB in urban areas.}
\label{linzclouds}
\end{figure}

Several factors may in principle account for the observed variation of the NSB throughout the year. Among them are the leaves of the trees, some of which absorb the light of streetlamps in cities during the summer months much more efficiently than during winter time. Alternatively, {snow cover and} the pre- to post-Christmas illumination could make winter skies brighter than summer skies, especially in cities. While these factors probably play a minor role, we found that the most decisive parameter governing the seasonal NSB variations are overcast skies. This is shown for Linz ({LGO}) in Fig.\ \ref{linzclouds} and is in accordance with previous studies, e.g.\ for the case of Berlin \citep{kyba2011cloud}.
{The basis of the okta cloud values for Linz are measurements with a partially automated weather station ('TAWES'). It is located at an altitude of 262m a.s.l. and at a distance of 1,73 km from the SQM at LGO. Measurements are done three times a day. Okta values, by definition, range from 0 to 8, where '0' means clompletely clear skies, while '8' denotes a completely overcast sky.}

For the city of Hong Kong, it has been shown by \citet{so2014} that if the cloud coverage increases from 40\% to 85\%, the urban sky brightens by about 2 magnitudes. This is consistent with our Fig.\ \ref{linzclouds}. Recently, \citet{jechow2017imaging} have shown that cloud coverage leads to a significant brightening of the night sky also far away from cities.

Most previously published papers, however, have focused on the seasonal NSB variations at remote observatories under \textit{clear skies}, since the telescopes used to measure the sky brightness are operated only under such conditions and astronomers are in general not interested in the brightness of the overcast sky. \citet{patat2008}, e.g., found a semi-annual oscillation (SAO) of the sky background brightness with a Johnson V amplitude of about 0.5 mag and with minima (darkness records) in July to August {(during southern hemisphere's winter)} and in December to January {(during southern hemisphere's summer)}. The SAO detected by Patat is more prominent at comparatively longer wavelengths (V, R, I) and almost insignificant at short wavelengths (Johnson B band). {Note that all of Patat's measurements were made in the southern hemisphere.}

\citet{liu2003}, by contrast, reported a different kind of seasonal variation of the clear night sky. At their observatory Xinglong Station, about 110 km NE of Beijing, they reported darker NSB values during fall to winter and brighter values during spring to summer. The authors presume that the darker skies in fall and winter are due to smaller scattering of ambient light and that, vice-versa, the brighter skies in spring and summer nights are due to enhanced scattering effects (with more desert dust particles in the air in summer). An amplitude of the seasonal variation is not given in their study.

\section{Correlation between night sky brightness and aerosol content}

For some of our NSB measurement stations in Upper Austria, simultaneously recorded data on the aerosol content are available. {They are based on measurements which are taken twice an hour. SQMs and particulate matter measuring devices are installed at one and the same location and essentially at the elevation (deviation $\le$2m).} We studied possible correlations between the NSB and the corresponding ``PM\,10 values'', which represent the concentration of particles with diameters between 2.5 and 10 $\mu$m in the atmosphere. Alternatively, the aerosol content can be expressed in terms of ``PM 2.5 values'', which refer to particles smaller than 2.5 $\mu$m. However, we found that in our case PM\,10 and PM\,2.5 are strongly correlated.

\begin{figure}[]
\centering
\includegraphics[width=1.0\textwidth]{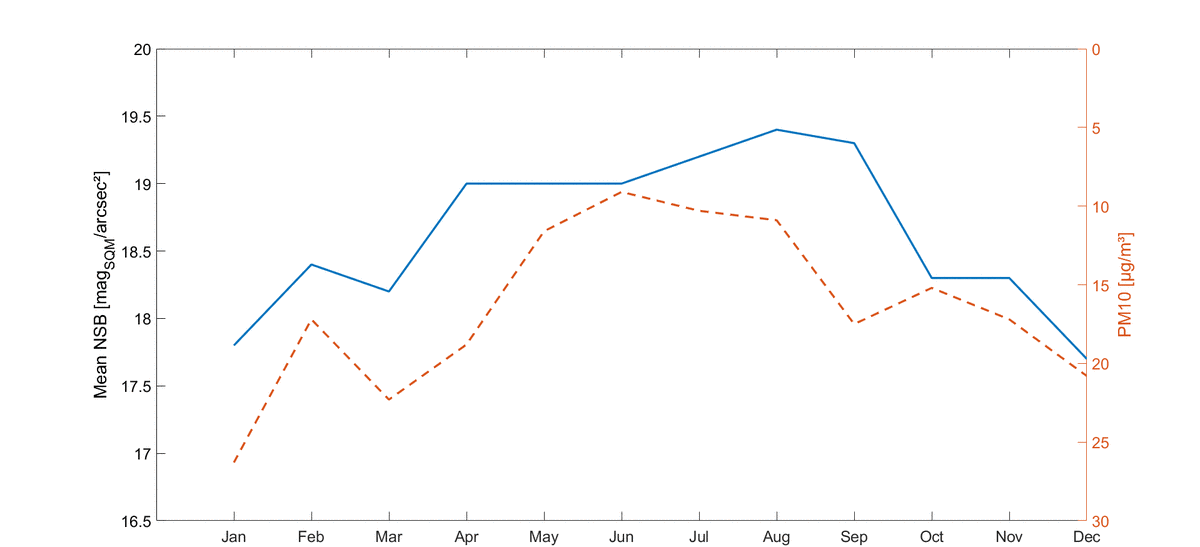}
\caption{Monthly averages of the NSB versus aerosol (PM\,10) concentration at the urban station VOE in 2016.}
\label{aerosols}
\end{figure}

As in the case of clouds, we first studied the relation between average NSB and average PM\,10 values, where ``averages'' were calculated on a monthly level (i.e.\ monthly averages of the $<$NSB$>$ values). Figure \ref{aerosols} shows the annual course of both quantities at the station of V\"ocklabruck ({VOE}) for the year 2016. 
A similar trend in average PM\,10 concentration and average NSB can be clearly seen in this case, with darker skies and lower PM\,10 values in summer to early autumn and brighter skies as well as higher PM\,10 values in winter. {Note that the plotted aerosol and NSB measurements are from the same location and elevation.}

\begin{figure}[]
\centering
\includegraphics[width=1.0\textwidth]{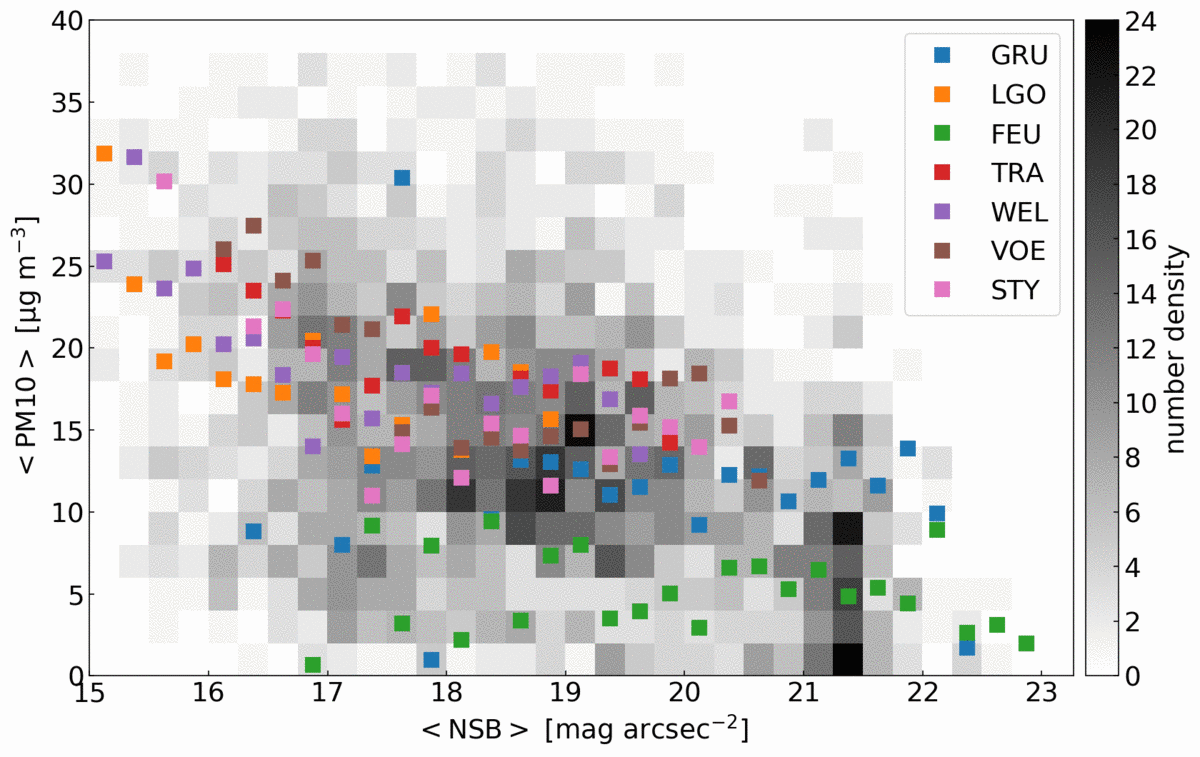}
\caption{Density plot of the aerosol content (PM\,10, daily average) and the nocturnal average NSB ($<$NSB$>$) values
for seven selected stations during the year 2016. The grayscale indicates the number density of points in the NSB-PM10-plane, 
while the overplotted smaller squares indicate the mean PM10 value per NSB bin for the different stations as colour-coded in the legend.}
\label{scatter}
\end{figure}

Furthermore, we generated a density plot of single-night mean NSB values versus available 24-hour-averages of the PM\,10 concentration for the year 2016 at seven selected stations (see Fig.\ \ref{scatter}).
As we recognize from this figure, there is a clustering of points between 21 and 22 mag$_{SQM}$/arcsec$^{2}$, which corresponds to clear nights in rural areas with low aerosol contents. Values beyond 22 mag$_{SQM}$/arcsec$^{2}$ only occur for PM\,10 $<$ 10 $\mu$g/m$^{3}$ and during overcast nights. For the urban areas and their surroundings (left half of Fig.\ \ref{scatter}), the overall scatter is obviously quite large, but still a trend of $<$NSB$>$ decreasing with PM\,10 can be seen.

{This confirms earlier results obtained by \citet{sciezor2014}, who also detected a maximum of the PM\,10 concentration during the winter months and hence an additional brightening of the night-sky due to enhanced scattering. While their data indicate that during December and January, the PM\,10 concentration is about 5 times higher than in August, our results suggest only a seasonal difference by a factor of 2-3, again for (sub)urban regions and low elevations. Considering the larger population of the Cracow metropolitan area and the high hibernal coal consumption in that region, this seems quite plausible. \'Sci\k{e}{\.z}or \& Kubala as well found a linear relationship between both quantities, PM\,10 and NSB -- which coincides with the results in our Fig.\ \ref{scatter}.}

{It should be noted that a previous Dutch study by \citet{denouter2011} had found no correlation between night sky brightness and aerosol content. However, their study was not based on monthly averages.}

\section{Correlation between NSB and population of the respective communities}

Since the seminal paper by \citet{walker1977effects}, we know that in countries and regions with similar degree of economic development, there is a positive correlation between the number of inhabitants of a community and its average NSB. 

While Walker showed that the population of a community and the respective lumen output are related to each other by a power law, we examine the correlation between population and the night sky brightness (i.e.\ the zenithal luminance) during clear, moonless nights.

\begin{figure}[]
\centering
\includegraphics[width=1.0\textwidth]{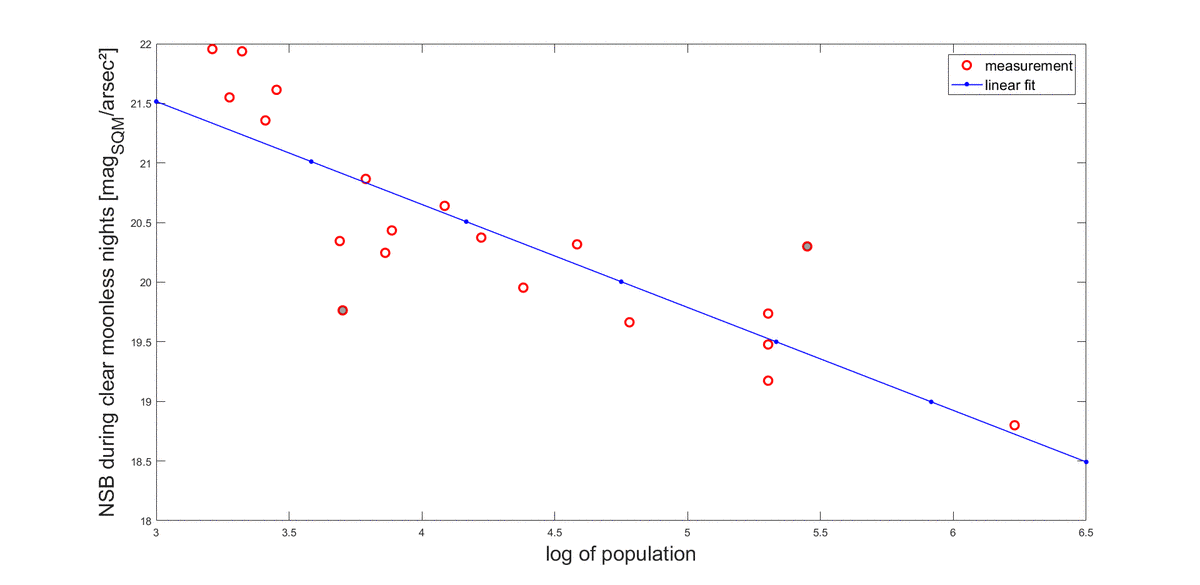}
\caption{Correlation between the NSB and the logarithm of the population for 20 monitoring stations in Upper Austria (excluding the stations in uninhabited areas), one at the outskirts of Graz ({GRA}) and one in Vienna ({IFA}) (bottom right). The solid line represents a linear fit of the data. Points above the solid line correspond to cities and villages where the sky is darker than expected or where the measurement station is clearly outside the city center. {Two stations, GRA and STW, are denoted with filled circles for reasons explained in the text. The line of best fit – excluding GRA and STW – corresponds to the equation: NSB = 24.4 mag - 0.91 log (P).}}
\label{population}
\end{figure}

As can be seen from Fig.\ \ref{population}, we do find a clear correlation between the decadic logarithm of the population (log(P)) and the clear-night-NSB (with a correlation coefficient of 0.85). Measurements from virtually uninhabited regions had to be omitted in this graph as no meaningful population could be assigned to them. Note that the town of Steyregg-Weih ({STW}) has a sky brightness that is entirely dominated by the nearby, much larger city of Linz ({e.g.\ LGO}). Obviously, the points below the solid line represent those communities which have brighter skies than expected based on their population, while the opposite is true for the points above the solid line -- among them are Graz ({GRA}) (at log (P) = 5.45) and Vienna ({IFA, bottom right data point, log (P) = 6.26)}. The three vertically aligned points at log (P) = 5.3 represent our three different measurement stations in Linz ({LSW, LSM, LGO}).

One should add that population is not the single factor determining the clear-sky NSB over a city or inhabited region. Rather, different lighting policies also have a strong influence on light pollution. {For example, the per capita lumen output varies quite strongly within Europe (not so much within single countries, however). Upward light emissions may correlate with the wealth of a region, but may also be influenced by local traditions of using or not using fully shielded lamps. Lighting curfews play an additional role in the abatement of light pollution.}


\section{Towards an analysis of long-term trends in the NSB (2012--2017)}

The question how the (mean) NSB at a given site is changing over the years is an important one both scientifically and practically, since responses may act as incitement for the abatement of light pollution.
However, when trying to determine long-term trends of the NSB, one is facing several challenges. 

1) Meteorological conditions may strongly vary from year to year. Examples include the number of overcast nights or the respective local duration of the snow coverage. These meteorological parameters can considerably influence the results of NSB measurements. 

2) A second problem, examined in detail by \citet{so2014}, arises from the aging of the SQM housing window (e.g.\ by solar UV irradiation). Based on visual examination, we have good reasons to assume that the loss of transparency of our housing windows is smaller than reported by So(2014) because of our higher latitudes and therefore lower solar elevations. Nevertheless, also at our latitudes, the transmission loss of the polycarbonate window  may still introduce a systematical apparent ``darkening'' of the night sky in the range of 0.1\, mag$_{SQM}$/arcsec$^{2}$ during the first year of operation, with a significant tendency to decrease afterwards \cite[Fig.\ 2.11]{so2014}. 
{For the longest series of measurements within our network, i.e.\ for the data from the Vienna University Observatory ({IFA}), which cover half a decade now, we also examined this apparent darkening effect by taking the 'jellyfish' plots and studying the shift towards the upper (darker) between 2012 and 2017. We found an apparent ``darkening'' of the darkest (= clearest) nights by about 0.15\, mag$_{SQM}$/arcsec$^{2}$, but within five years. This effect is shown in the Appendix, Fig.\ A4}.

3) Thirdly, the current transition towards solid state lighting often leads to a significant change in the spectrum and overall colour of the skyglow that we measure. For example, a large-scale replacement of high pressure sodium lamps with LEDs of 4000\,K colour temperature leads to a strong shift of the perceived colour of the night sky towards blue wavelengths. SQM measurements of the NSB in a city that undergoes such a transition would lead to the (partially wrong) conclusion of an only small increase (0.24 mag) in the luminance of the night sky. By contrast, the scotopic human vision might perceive the same change as an increase in the sky's luminance by about one full magnitude (see \citet{sanchez2017sky}, esp.\ Tab.\ 1). This means that based on SQM measurements alone, we may strongly underestimate the growth rates of light pollution.

4) A fourth challenge is the fact that the annual change in the NSB -- for most developed countries and cities -- amounts to less than 0.05 magnitudes per year \citep{narisada2013light}.
This change is of the same order of magnitude as the accuracy of SQM measurements and therefore only measurable after more than two years, which in turn is about the same time span as the one covered by most of our data.

For the above mentioned site ({IFA}), we additionally applied a new data reduction routine and calibrated the zenithal NSB values obtained in clear, moonless nights with the 'Skycal' model of the European Southern Observatory \citep{Noll+2012,Jones+2013}. Only observing times during which the Sun was at least 15$^{\circ}$ below the horizon and when the course of the NSB was smooth and flat were used for this in-depth analysis, the details of which will be explained in a forthcoming paper (Puschnig et al., in preparation).
Here is a short summary of our results for the development of the calibrated clear-sky-NSB at IFA (recall that this site is at 3\,km distance from the city center):

\begin{itemize}
    \item{The analysis was performed for the time 2012-03-01 to 2017-07-01.}
    \item{The average (2012--2017) clear-sky-NSB at the site was 2.73\,mag$_{SQM}$/arcsec$^2$ above the natural zenithal NSB, i.e.\ 12 times as much as the natural value.}
    \item{The change of the clear-sky-NSB during this time was a (hardly significant) darkening by 0.018$\pm$0.005 \,mag$_{SQM}$/arcsec$^2$/yr.}
\end{itemize}

This would imply a \textit{decrease}\/ in light pollution by 1.8\% at the site Vienna University Observatory ({IFA}). We are still doubtful about this, since it would mean that the noticeable efforts, during the past few years, to reduce light pollution from public illumination, have indeed outweighed the evident simultaneous increase in commercial lighting. Further efforts to study the long-term development of the NSB and to derive robust trends will be necessary.

{Based on calibrated satellite data for the years 2012-2016, \citet{Kybae1701528} found that there is an increase in the Earth‘s artificially lit outdoor areas by 2.2 \% per year and a total total radiance growth rate which has approximately the same value. National light pollution growth rates, according to Kyba et al., are difficult to determine, but some seem to be much larger than 2 \%. They roughly correlate with the national gross domestic product. Only in very few countries, a \textit{decrease} in artificial outdoor lighting (with respect to intensity and/or area) is found.}

\section{Conclusions}

For the first time, we monitored the night sky brightness with all kinds of meteorological as well as moonlight conditions at 26 stations in Eastern Austria, with a focus on the county of Upper Austria. More than 10 million data points of NSB measurements were obtained during 2015 and 2016, typically with a sampling rate of 0.0167 Hz.

For the resulting large datasets, we developed several ways to display them and to derive criteria on the local quality of the (zenithal) night sky. The presented tools and criteria include:

\begin{itemize}
    \item{mean NSB histograms: these often show a bimodal frequency distribution for urban locations (with a 'bright nights' peak for overcast conditions and a 'dark nights' peak for clear, moonless skies) versus an unimodal frequency distribution for rural locations, where clouds do not strongly enhance the NSB.}
    \item{circalunar mean-nocturnal-NSB-variation: these plots shows a distinctive periodicity for locations like Krippenstein ({KRI}), Feuerkogel ({FEU}), locations in/near the National Park 'Kalkalpen' (BOD, ZOE, LOS) and others, while a stochastic, weather-dominated pattern emerges for urban and metropolitan areas (see Appendix, Fig.\ \ref{circalunar1}-\ref{circalunar2}). Circalunar $<$NSB$>$ periodicity should more frequently be used for constraining the light pollution at a given site, since this criterion delivers clear and chronobiologically relevant information. Fourier periodograms will be helpful for further studies in this field.}
    \item{Annual NSB variation, visualized by 'hourglass' plots: by this means again, the circalunar periodicity or non-periodicity is very intuitively presented, but the annual variation of the night lengths and NSB also become evident, as do the potential NSB variations within individual nights. 'Near-natural' patterns of the hourglass plots emerge for remote alpine locations (see Fig.\ \ref{hourglass2016_1}).}
    \item{Density ('jellyfish') plots: the bimodality mentioned for the 'urban' mean NSB histograms, namely between bright overcast and clear dark nights emerges in the form of two vertically separated high-density-clusters in these diagrams. The nocturnal gradient of NSB (t) can be visualized in these plots: see Fig.\ \ref{jellyfish} and Figs.\ \ref{jelly1}-\ref{jelly2}.}
    \item{Averaged jellyfish plots: In these diagrams, introduced in Sect.\ 4.5, each location is characterized by a single (long-time-averaged) NSB (t) graph. The upper envelope each such graph delivers NSB$_{avg, max}$, a robust measure of the maximum darkness reached at a site on average, i.e.\ with \textit{all} meteorological conditions. Our NSB$_{avg, max}$ values for Upper Austria range from 17.7 to 21.1 mag$_{SQM}$/arcsec$^2$, while the respective darkness records for clear nights amount to 19.0 to 21.9 mag$_{SQM}$/arcsec$^2$. Furthermore, average nocturnal gradients during the core hours of darkness, 't-grad' (10~PM -- 2~AM local time), may be derived. For these we find values between 0.0 and 0.14\,mag$_{SQM}$ per hour.}
    \item{Seasonal $<$NSB$>$ variations: these are, if we include all weather conditions again, much stronger at urban locations, with brighter winter nights and darker summer nights. This trend is the opposite of the natural seasonal variation of darkness, but in accordance with the higher number of cloud-free nights in Central European summers as compared to winters.} 
    \item{NSB and aerosol content: Like for the cloud okta values, we found a co-variation of the monthly mean $<$NSB$>$ values with PM\,10 values at least for some of our stations.}
    \item{NSB and population: Expectedly, the clear-night NSBs do correlate with corresponding populations, but there are slight deviations for cities and communities that have a smaller lumen-per-capita output than others.}
\end{itemize}

In the light of our results, we conclude that the protection of those sites where we still find darkest NSB values close to 21.8 mag$_{SQM}$/arcsec$^2$ should be urgently envisaged. This is the case for Krippenstein ({KRI}) close to the Dachstein plateau, Mount Feuerkogel ({FEU}), and three places in and near the National Park 'Kalkalpen'. This protected area has a size of 208\,km$^2$. Including some of the adjacent communities, a Dark Sky Reserve more than 700\,km$^2$ in size should be established and could potentially reach the 'gold tier' of the International Dark Sky Association. {For a first drafted boundary line of such a reserve, we refer back to Fig.\ \ref{geodistribution}, red area}.


\section*{Acknowledgements}

We gratefully acknowledge extensive work done by Heribert Kaineder and Martin Waslmaier -- both with the provincial government of Upper Austria -- who spent a lot of time with the setup of 23 SQM measurement sites. We also acknowledge financial support by their department. 
Salvador Ribas and his collaborators kindly provided a python code which served as starting point for the development of the codes we used to generate our hourglass plots.
Armin Luntzer, Vienna, helped us a lot with data processing and with the preparation of some figures. Anthony Tekatch from Unihedron was very supportive in answering questions on the SQM.
As for the measurement station at Graz-Lustb\"uhel, we thank Robert Greimel for setting up the instrument and providing us with the recorded data. Furthermore we thank G\"unther Wuchterl for fruitful discussions on the whole topic of this paper.
Two anonymous referees contributed significantly to the improvement of the original paper version.

\pagebreak

\bibliography{ref}

\bibliographystyle{unsrtnat}

\pagebreak

\section*{Appendix}

\setcounter{figure}{0}
\makeatletter 
\renewcommand{\thefigure}{A.\@arabic\c@figure}
\makeatother

In the following, we present a complete set of NSB histograms, circalunar plots and density ('jellyfish') plots for all 23 measuring stations in Upper Austria, always referring to the year 2016. All figures are arranged by increasing average NSB, i.e.\ in the same way as Tab.\ \ref{nsbmax-t} and Fig.\ \ref{hourglass2016_1}-\ref{hourglass2016_2}. 
For a detailed description of the different plot types see above, Sect.\ 4.

The individual plots for all stations and for several years can be found at \\ \url{http://www.univie.ac.at/nightsky/} in sub-directories with the corresponding names, i.e.\ 'circalunar', 'histograms', 'hourglass' and 'jellyfish'.

\begin{figure}[]
\centering
\includegraphics[width=1.0\textwidth]{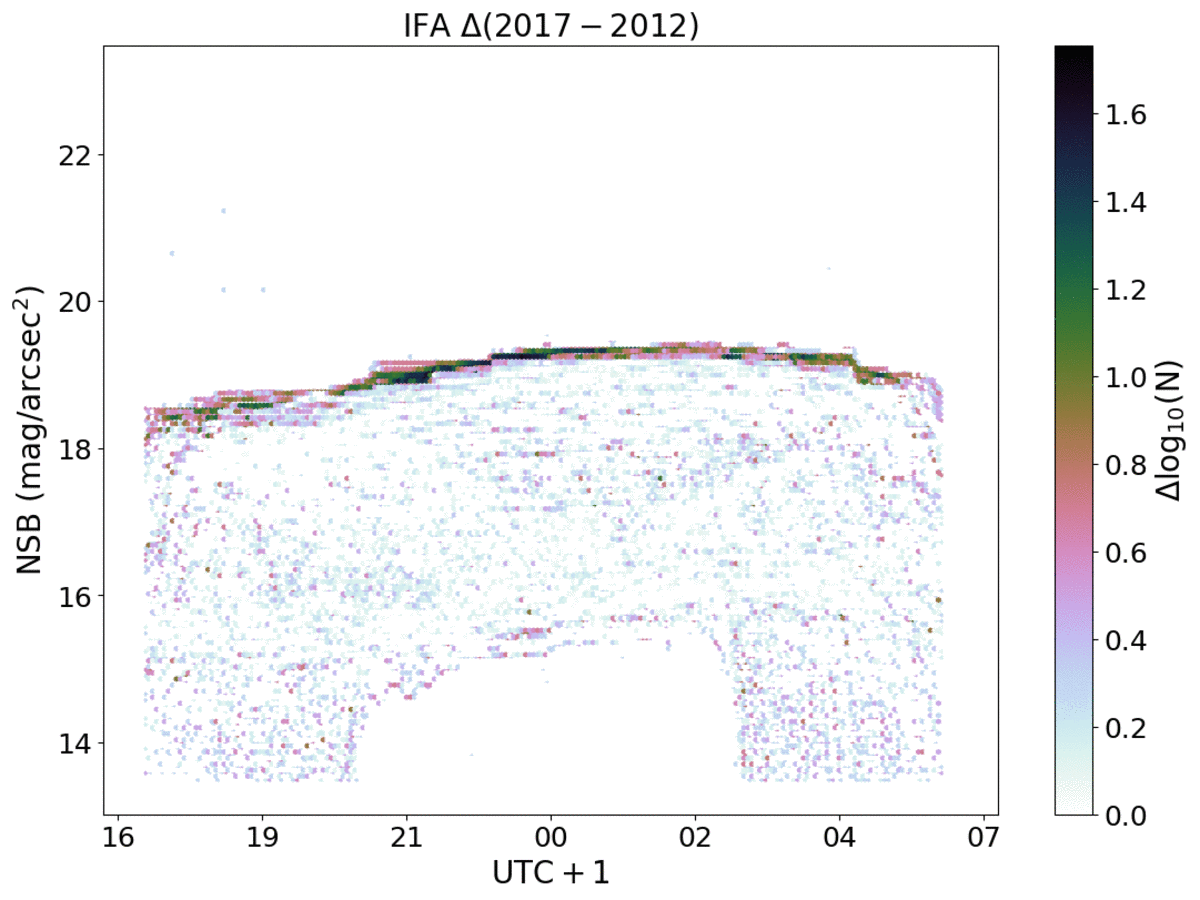}
\caption{Difference between the density plots of the NSB measurements performed at the Vienna University Observatory (IFA) in 2017 and in 2012. From this diagram, we infer a shift towards darker values by about 0.15\,mag$_{SQM}$/arcsec$^{2}$ within five years. However, we take this is only as an apparent indication of a darkening of the night sky at IFA (see Sect.\ 8 for more details).}
\label{delta-jelly}
\end{figure}

In addition, Fig.\ \ref{delta-jelly} shows the difference between the NSB densities recorded at the station IFA in 2017 and 2012 as discussed in Sect.\ 8.

\begin{subfigures}
\begin{figure}[h]\centering
  \includegraphics[width=1.0\textwidth]{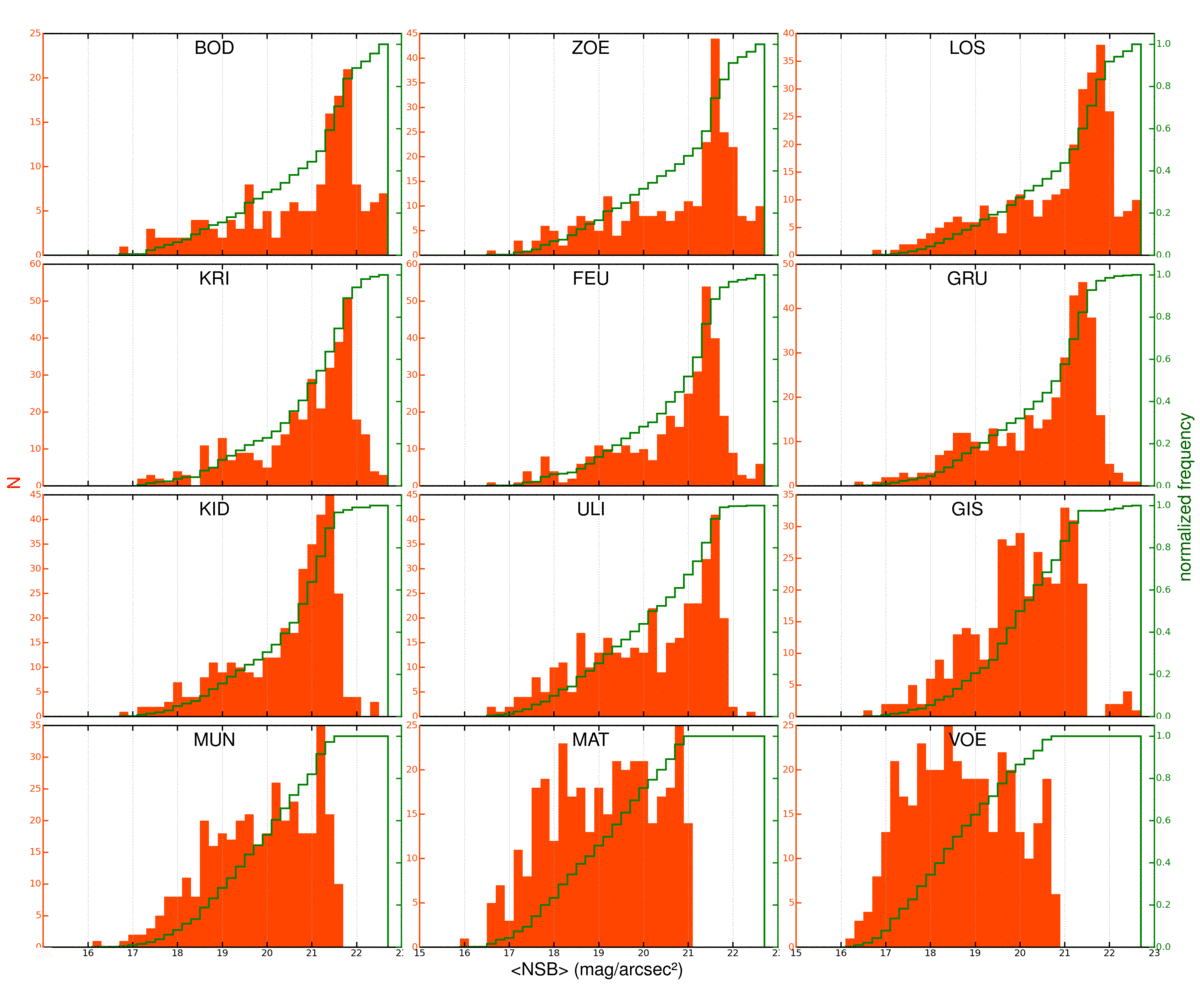}
  \caption{\label{hist1}Histograms of the mean nocturnal NSB in 2016 at 23 different locations in Upper Austria arranged by increasing average NSB (part 1: BOD -- VOE). Note the peak at the rural sites between 21 and 22, which represents the NSB value for clear nights. } 
\end{figure}
\begin{figure}\centering
  \includegraphics[width=1.0\textwidth]{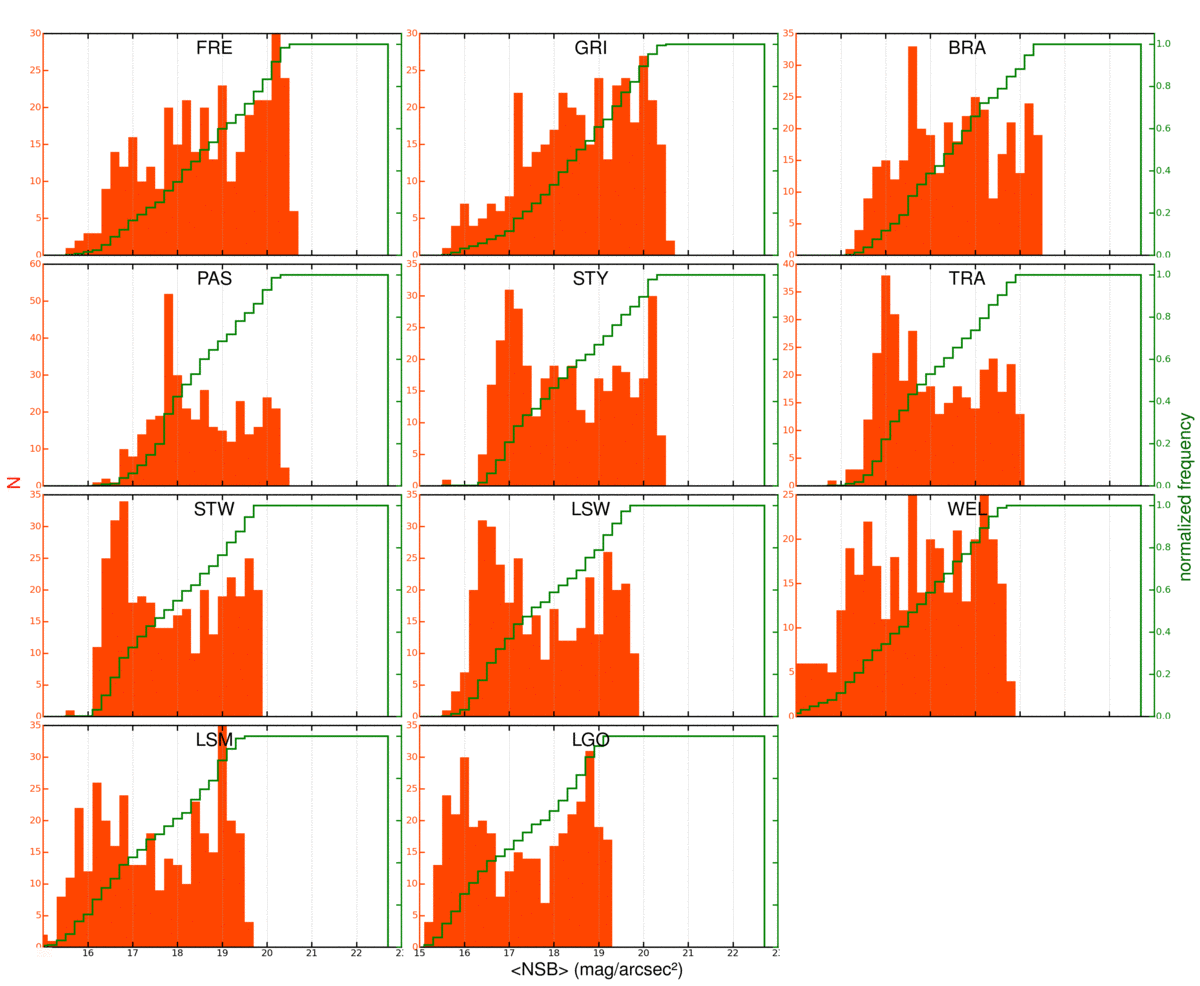}
  \caption{\label{hist2}Histograms of the mean nocturnal NSB in 2016 at 23 different locations in Upper Austria arranged by increasing average NSB (part 2: FRE -- LGO). With increasing population a second peak emerges, which seems to represent overcast nights.} 
\end{figure}
\end{subfigures}

\begin{subfigures}
\begin{figure}\centering
  \includegraphics[width=1.0\textwidth]{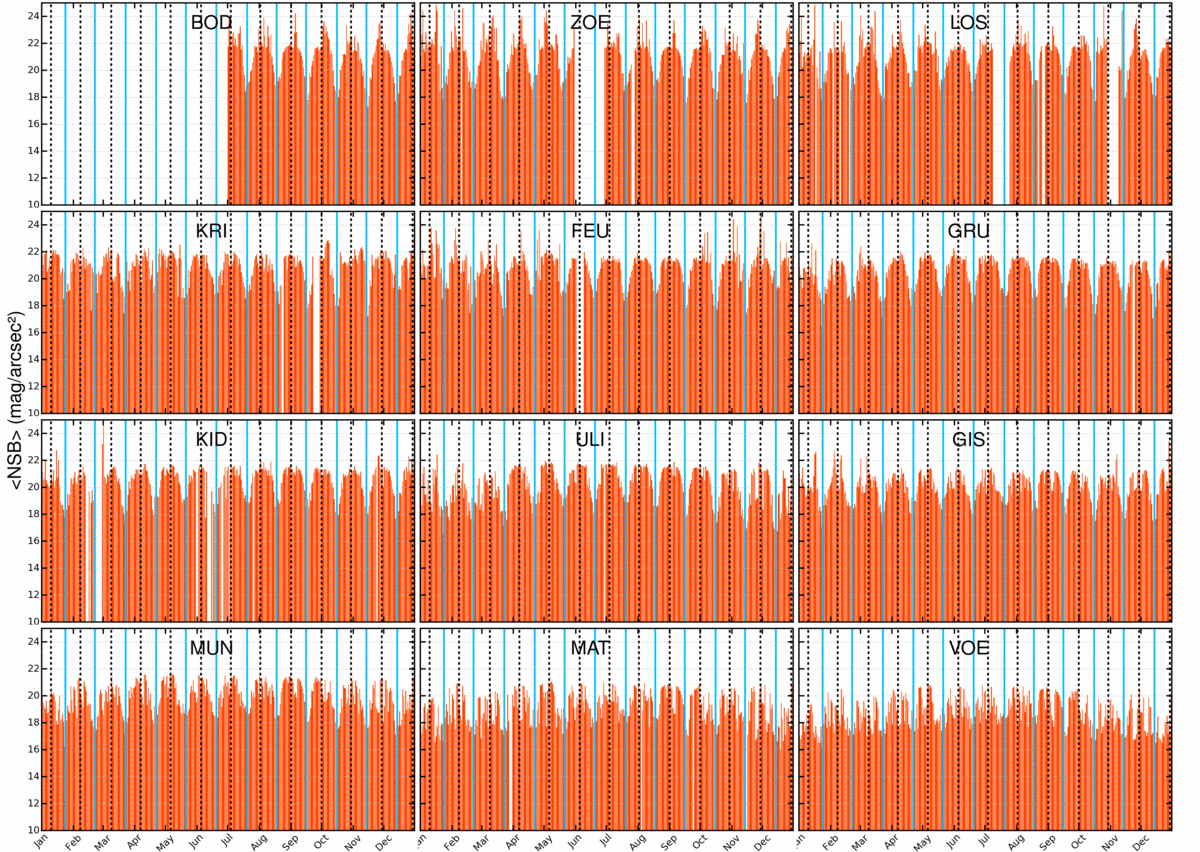}
  \caption{\label{circalunar1}Circalunar bar charts of the mean nocturnal NSB in 2016 at 23 different locations in Upper Austria (part 1: BOD -- VOE). Dotted vertical lines denote the times of New Moon. Note the gradual transition from very pronounced to poor circalunar periodicity of the $<$NSB$>$.} 
\end{figure}
\begin{figure}\centering
  \includegraphics[width=1.0\textwidth]{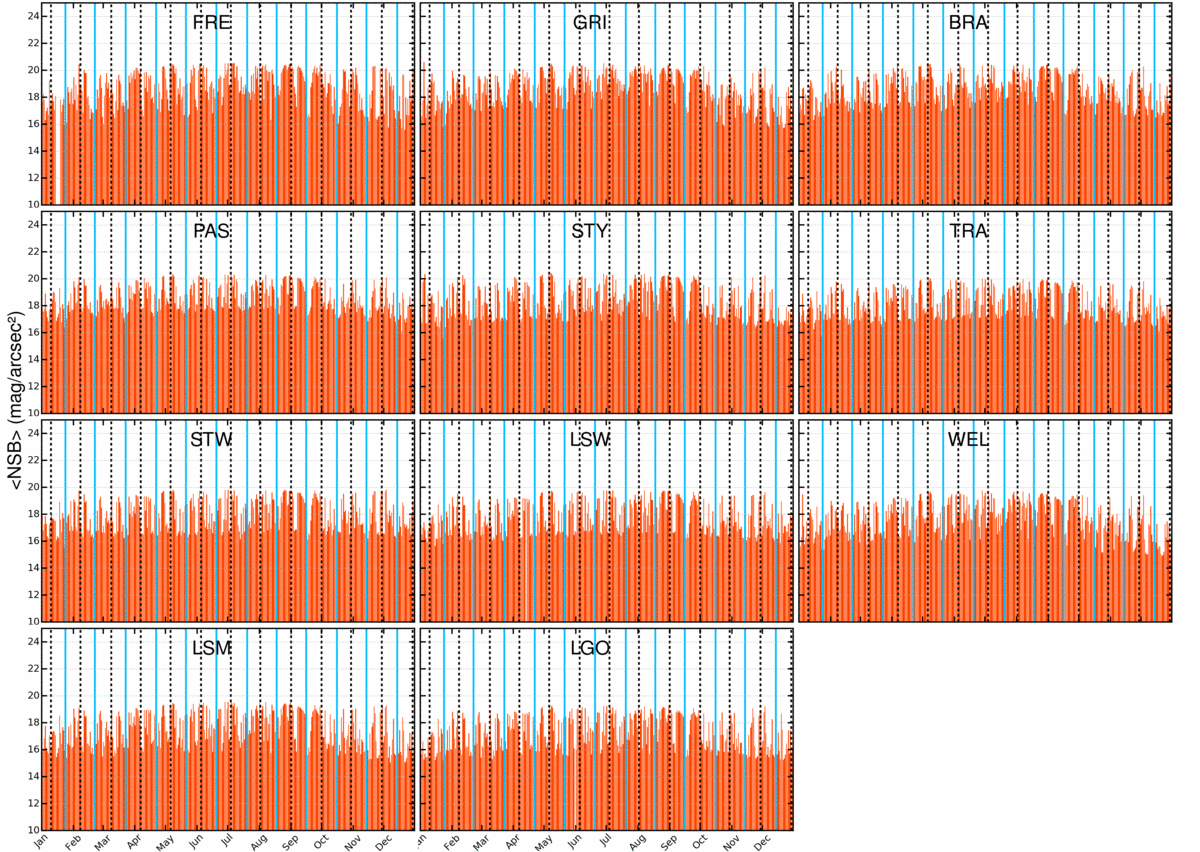}
  \caption{\label{circalunar2}Circalunar bar charts of the mean nocturnal NSB in 2016 at 23 different locations in Upper Austria (part 2: FRE -- LGO). The locations covered in this figure are characterized by increasingly non-periodic patterns} 
\end{figure}
\end{subfigures}

\begin{subfigures}
\begin{figure}\centering
  \includegraphics[width=1.0\textwidth]{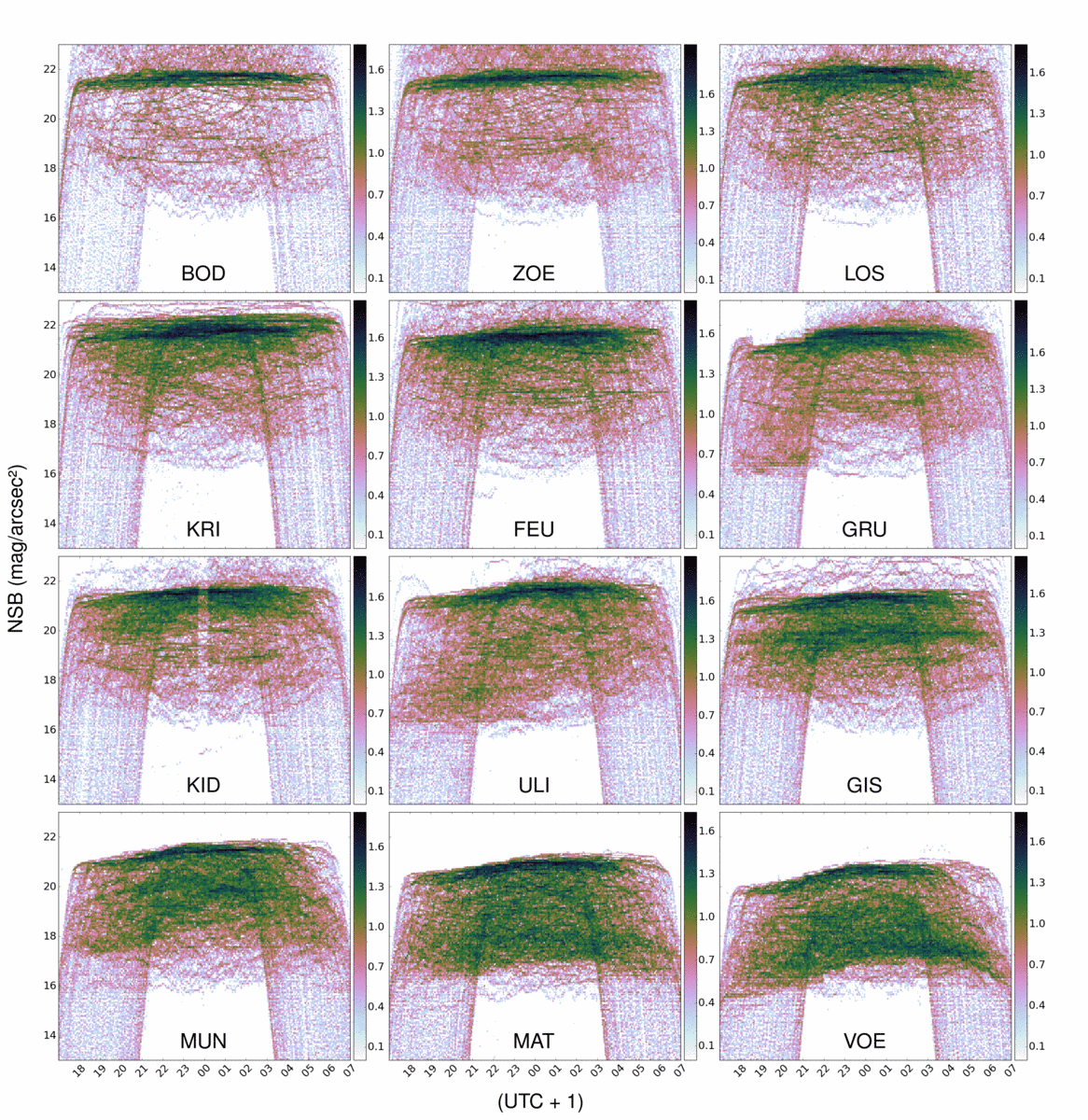}
  \caption{\label{jelly1}Jellyfish (density) plots of the NSB values in 2016 at 23 different locations in Upper Austria (part 1: BOD -- VOE).} 
\end{figure}
\begin{figure}\centering
  \includegraphics[width=1.0\textwidth]{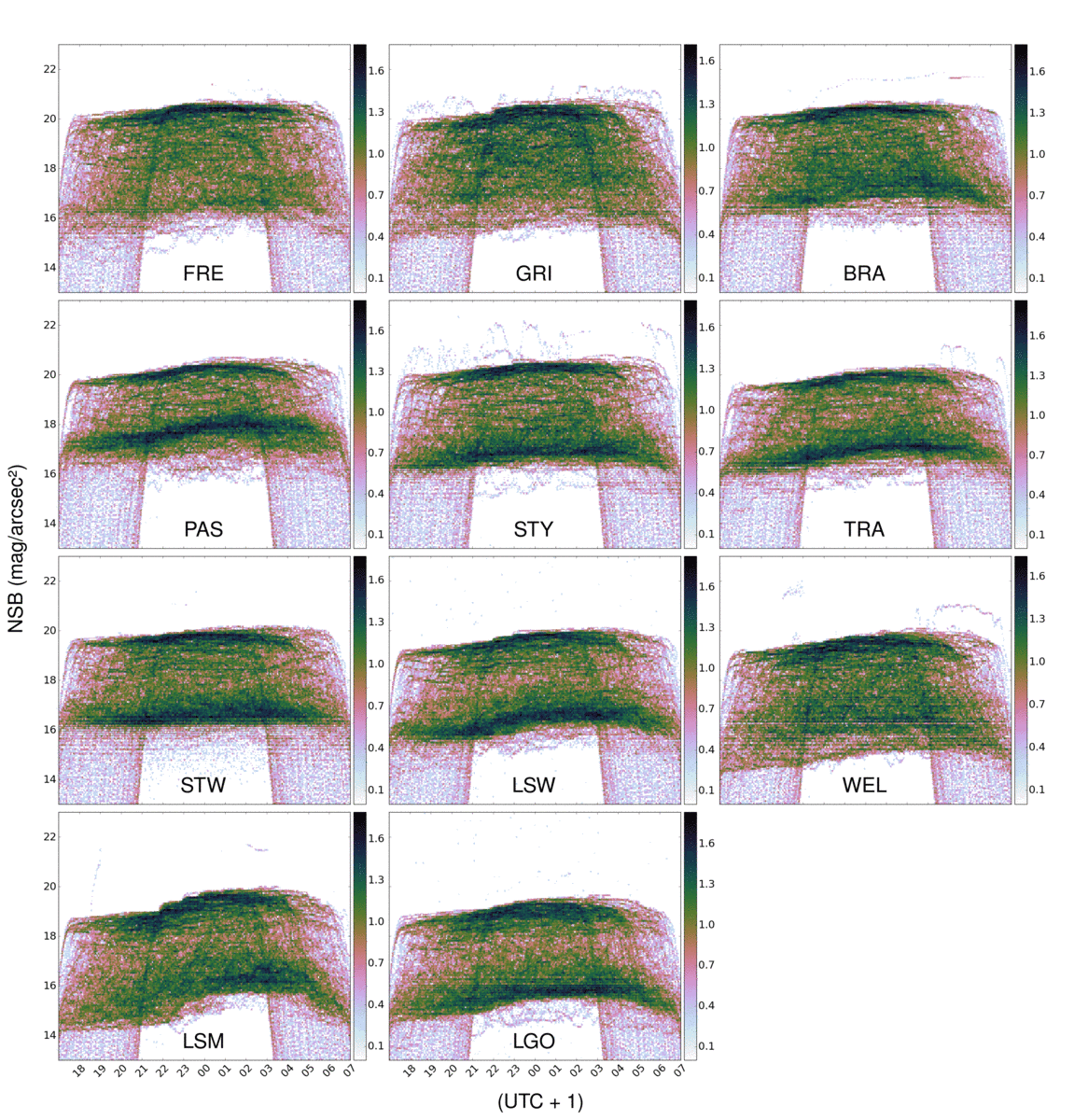}
  \caption{\label{jelly2}Jellyfish (density) plots of the NSB values in 2016 at 23 different locations in Upper Austria (part 2: FRE -- LGO).} 
\end{figure}
\end{subfigures}

\nolinenumbers

\end{document}